\documentclass[manuscript=article,layout=traditional]{achemso}

\usepackage{xcolor} 
\usepackage[noend]{algpseudocode}
\usepackage{array}
\usepackage{amsfonts}
\usepackage{graphicx}
\usepackage{bm,mathrsfs}
\usepackage{epstopdf}
\usepackage{amssymb}
\usepackage{tikz,pgfplots}
\usetikzlibrary{decorations,arrows,shapes,trees}
\usepackage{tikz-3dplot}
\usepackage{amssymb,amsmath,amsthm}
\usepackage{amsopn}
\usepackage{listings}
\usepackage{adjustbox}
\usepackage{longtable}
\usepackage{multirow}
\usepackage{hyperref}
\usepackage{amssymb}
\usepackage{pgfplots}
\usepackage{pgfplotstable}
\usepackage{grffile}
\usepackage{siunitx}
\usetikzlibrary{arrows,shapes,plotmarks}
\usetikzlibrary{decorations.pathreplacing}
\usepgfplotslibrary{groupplots}
\usetikzlibrary{matrix}
\usepackage[]{changes}
\newcommand{\cref}[2]{\hyperref[#2]{#1~\ref*{#2}}}
\newcommand{\figref}[1]{\hyperref[#1]{Fig.~\ref*{#1}}}
\newcommand{\secref}[1]{\hyperref[#1]{Section~\ref*{#1}}}
\newcommand{\tabref}[1]{\hyperref[#1]{Table~\ref*{#1}}}
\newcommand{\eqnref}[1]{\hyperref[#1]{Eq.~(\ref*{#1})}}
\newcommand{\Algref}[1]{\hyperref[#1]{Algorithm~\ref*{#1}}}
\newcommand{\Tabref}[1]{\hyperref[#1]{Table~\ref*{#1}}}
\hypersetup{
	colorlinks,
	linkcolor={blue!50!black},
	citecolor={blue!50!black},
	urlcolor={blue!80!black},
	anchorcolor = {blue!80!black},
	filecolor = {blue!80!black},
	menucolor = {blue!80!black},
	runcolor = {blue!80!black}
}

\pgfplotsset{compat=1.8}
\usepackage{soul}
\usepackage{rotating}
\usepackage{url}
\usepackage{algorithm}
\usepackage{enumitem}
\usepackage{subcaption}
\newcommand{\Vector}[1]{\underline{\mathbf{#1}}}
\newcommand{\UnitVector}[1]{\underline{\mathbf{\hat{#1}}}}
\newcommand{\Tensor}[1]{\underline{\underline{\mathbf{#1}}}}

\newdimen\HilbertLastX
\newdimen\HilbertLastY
\newcounter{HilbertOrder}

\def\DrawToNext#1#2{%
  \advance \HilbertLastX by #1
  \advance \HilbertLastY by #2
  \pgfpathlineto{\pgfqpoint{\HilbertLastX}{\HilbertLastY}}
}

\def\Hilbert[#1,#2,#3,#4,#5,#6,#7,#8] {
  \ifnum\value{HilbertOrder} > 0%
  \addtocounter{HilbertOrder}{-1}
  \Hilbert[#5,#6,#7,#8,#1,#2,#3,#4]
  \DrawToNext {#1} {#2}
  \Hilbert[#1,#2,#3,#4,#5,#6,#7,#8]
  \DrawToNext {#5} {#6}
  \Hilbert[#1,#2,#3,#4,#5,#6,#7,#8]
  \DrawToNext {#3} {#4}
  \Hilbert[#7,#8,#5,#6,#3,#4,#1,#2]
  \addtocounter{HilbertOrder}{1}
  \fi
}

\def\hilbert((#1,#2),#3){%
  \advance \HilbertLastX by #1
  \advance \HilbertLastY by #2
  \pgfpathmoveto{\pgfqpoint{\HilbertLastX}{\HilbertLastY}}
  \setcounter{HilbertOrder}{#3}
  \Hilbert[1mm,0mm,-1mm,0mm,0mm,1mm,0mm,-1mm]
  \pgfusepath{stroke}%
}

\ifpdf
  \DeclareGraphicsExtensions{.eps,.pdf,.png,.jpg}
\else
  \DeclareGraphicsExtensions{.eps}
\fi

\emergencystretch=\maxdimen
\definecolor{cpu3}{HTML}{F44336}
\definecolor{cpu4}{HTML}{2196F3}
\definecolor{cpu1}{HTML}{4CAF50}
\definecolor{cpu2}{HTML}{FFC107}
\definecolor{gpu3}{HTML}{EF9A9A}
\definecolor{gpu4}{HTML}{90CAF9}
\definecolor{gpu1}{HTML}{A5D6A7}
\definecolor{gpu2}{HTML}{FFE082}

\definecolor{cpu5}{HTML}{9932CC}

\definecolor{sq_b1}{RGB}{37,52,148}
\definecolor{sq_b2}{RGB}{44,127,184}
\definecolor{sq_b3}{RGB}{65,182,196}
\definecolor{sq_b4}{RGB}{127,205,187}
\definecolor{sq_b5}{RGB}{199,233,180}
\definecolor{sq_b6}{RGB}{255,255,204}

\definecolor{sq_r1}{RGB}{189,0,38}
\definecolor{sq_r2}{RGB}{240,59,32}
\definecolor{sq_r3}{RGB}{253,141,60}
\definecolor{sq_r4}{RGB}{254,178,76}
\definecolor{sq_r5}{RGB}{254,217,118}
\definecolor{sq_r6}{RGB}{255,255,178}

\definecolor{sq_g1}{RGB}{0,104,55}
\definecolor{sq_g2}{RGB}{49,163,84}
\definecolor{sq_g3}{RGB}{120,198,121}
\definecolor{sq_g4}{RGB}{173,221,142}
\definecolor{sq_g5}{RGB}{217,240,163}
\definecolor{sq_g6}{RGB}{255,255,204}

\definecolor{div_c1}{RGB}{230,171,2}
\definecolor{div_c2}{RGB}{102,166,30}
\definecolor{div_c3}{RGB}{231,41,138}
\definecolor{div_c4}{RGB}{117,112,179}
\definecolor{div_c5}{RGB}{217,95,2}
\definecolor{div_c6}{RGB}{27,158,119}
\definecolor{div_c7}{RGB}{215,48,39}

\definecolor{div_d1}{RGB}{215,25,28}
\definecolor{div_d2}{RGB}{253,174,97}
\definecolor{div_d3}{RGB}{255,255,191}
\definecolor{div_d4}{RGB}{171,217,233}
\definecolor{div_d5}{RGB}{44,123,182}

\definecolor{lineclr}{RGB}{0,0,0}
\definecolor{utorange}{RGB}{0,0,255}
\definecolor{utsecblue}{RGB}{255,255,0}
\definecolor{utsecgreen}{RGB}{255,0,0}
\definecolor{red!15}{RGB}{0,255,255}
\definecolor{fillclr5}{RGB}{0,255,0}
\definecolor{fillclr6}{RGB}{255,0,255}
\definecolor{fillclr7}{RGB}{255,255,255}
\definecolor{fillclr8}{RGB}{0,0,0}

\def\drawcubeI(#1,#2,#3,#4,#5){ % x,y,z,sz,line color
\coordinate (O) at (#1,#2,#3);
\coordinate (A) at (#1,#2+#4,#3);
\coordinate (B) at (#1,#2+#4,#3+#4);
\coordinate (C) at (#1,#2,#3+#4);
\coordinate (D) at (#1+#4,#2,#3);
\coordinate (E) at (#1+#4,#2+#4,#3);
\coordinate (F) at (#1+#4,#2+#4,#3+#4);
\coordinate (G) at (#1+#4,#2,#3+#4);
\draw[#5] (O) -- (C) -- (G) -- (D) -- cycle;% Bottom Face
\draw[#5] (O) -- (A) -- (E) -- (D) -- cycle;% Back Face
\draw[#5] (O) -- (A) -- (B) -- (C) -- cycle;% Left Face
\draw[#5] (D) -- (E) -- (F) -- (G) -- cycle;% Right Face
\draw[#5] (C) -- (B) -- (F) -- (G) -- cycle;% Front Face
\draw[#5] (A) -- (B) -- (F) -- (E) -- cycle;% Top Face
}

\def\drawcubeII(#1,#2,#3,#4,#5,#6,#7){ % x,y,z,sz,line color,fill color,opacity
\coordinate (O) at (#1,#2,#3);
\coordinate (A) at (#1,#2+#4,#3);
\coordinate (B) at (#1,#2+#4,#3+#4);
\coordinate (C) at (#1,#2,#3+#4);
\coordinate (D) at (#1+#4,#2,#3);
\coordinate (E) at (#1+#4,#2+#4,#3);
\coordinate (F) at (#1+#4,#2+#4,#3+#4);
\coordinate (G) at (#1+#4,#2,#3+#4);
\draw[#5,fill=#6,opacity=#7] (O) -- (C) -- (G) -- (D) -- cycle;% Bottom Face
\draw[#5,fill=#6,opacity=#7] (O) -- (A) -- (E) -- (D) -- cycle;% Back Face
\draw[#5,fill=#6,opacity=#7] (O) -- (A) -- (B) -- (C) -- cycle;% Left Face
\draw[#5,fill=#6,opacity=#7] (D) -- (E) -- (F) -- (G) -- cycle;% Right Face
\draw[#5,fill=#6,opacity=#7] (C) -- (B) -- (F) -- (G) -- cycle;% Front Face
\draw[#5,fill=#6,opacity=#7] (A) -- (B) -- (F) -- (E) -- cycle;% Top Face
}

\def\drawNodes(#1,#2,#3,#4,#5,#6,#7){ % x_min,x_max,y_min,y_max,z_min,z_max,min+stepSz
\foreach \x in {#1,#7,...,#2}{
	\foreach \y in {#3,#7,...,#4}{
		\foreach \z in {#5,#7,...,#6}{
				\draw[fill=red!60] (\x,\y,\z) circle (0.15);
				}
			}
	}				
		
}

\pgfplotsset{
  log x ticks with fixed point/.style={
      xticklabel={
        \pgfkeys{/pgf/fpu=true}
        \pgfmathparse{exp(\tick)}%
        \pgfmathprintnumber[fixed relative, precision=3]{\pgfmathresult}
        \pgfkeys{/pgf/fpu=false}
      }
  },
  log y ticks with fixed point/.style={
      yticklabel={
        \pgfkeys{/pgf/fpu=true}
        \pgfmathparse{exp(\tick)}%
        \pgfmathprintnumber[fixed relative, precision=3]{\pgfmathresult}
        \pgfkeys{/pgf/fpu=false}
      }
  }
}

\makeatletter
\newcommand\resetstackedplots{
\makeatletter
\pgfplots@stacked@isfirstplottrue
\makeatother
\addplot [forget plot,draw=none] coordinates{(48,0) (96,0) (192,0) (384,0) (768,0) (1536,0) (3072,0) (6144,0)};
}
\makeatother

\makeatletter
\newcommand\resetstackedplotsOne{
\makeatletter
\pgfplots@stacked@isfirstplottrue
\makeatother
\addplot [forget plot,draw=none] coordinates{(384,0) (768,0) (1536,0) (3072,0) (6144,0)};
}
\makeatother

\makeatletter
\newcommand\resetstackedplotsTwo{
\makeatletter
\pgfplots@stacked@isfirstplottrue
\makeatother
\addplot [forget plot,draw=none] coordinates{(16,0) (32,0) (64,0) (128,0) (256,0) (512,0) (1024,0) (2048,0) (4096,0) (8192,0) (16384,0) (32768,0)};
}
\makeatother

\makeatletter
\newcommand\resetstackedplotsThree{
\makeatletter
\pgfplots@stacked@isfirstplottrue
\makeatother
\addplot [forget plot,draw=none] coordinates{(2,0) (4,0) (8,0) (16,0) (32,0) (64,0)};
}
\makeatother

\makeatletter
\newcommand\resetstackedplotsFour{
\makeatletter
\pgfplots@stacked@isfirstplottrue
\makeatother
\addplot [forget plot,draw=none] coordinates{(4,0) (8,0) (16,0) (32,0) (64,0)};
}
\makeatother

\makeatletter
\newcommand\resetstackedplotsFive{
\makeatletter
\pgfplots@stacked@isfirstplottrue
\makeatother
\addplot [forget plot,draw=none] coordinates{(1,0) (2,0) (4,0) (8,0) (16,0) (32,0) (64,0) (128,0)};
}
\makeatother

\makeatletter
\newcommand\resetstackedplotsSix{
\makeatletter
\pgfplots@stacked@isfirstplottrue
\makeatother
\addplot [forget plot,draw=none] coordinates{(2,0) (4,0) (8,0) (16,0) (32,0) (64,0) (128,0)};
}
\makeatother

\definecolor{armygreen}{rgb}{0.29, 0.33, 0.13}
\definecolor{aurometalsaurus}{rgb}{0.43, 0.5, 0.5}
\definecolor{applegreen}{rgb}{0.55, 0.71, 0.0}
\definecolor{darkgreen}{rgb}{0.0, 0.4, 0.25}

 \usepackage{tgpagella}
\newtheorem*{remark}{Remark}
\usepackage{dirtytalk}
\newcommand{\prsoxs}{P-RSoXS}
\usepackage{tcolorbox}
\tcbuselibrary{theorems}
\usepackage{varwidth}
\tcbuselibrary{skins}
\tcbuselibrary{breakable}
\tcbsetforeverylayer{shield externalize}
\usepackage{titlecaps, cleveref}
\usetikzlibrary{decorations.pathreplacing,calligraphy}
\SectionNumbersOn
\title{CyRSoXS: A GPU-accelerated virtual instrument for Polarized Resonant Soft X-ray Scattering (\prsoxs)}

\author{Kumar Saurabh}
\affiliation{Department of Mechanical Engineering, Iowa State University, Ames, IA 50010, USA}
\author{Peter J. Dudenas}
\affiliation[nist]{Material Measurement Laboratory, National Institute of Standards and Technology (NIST), Gaithersburg, MD 20899, USA}
\author{Eliot Gann}
\affiliation[nist]{Material Measurement Laboratory, National Institute of Standards and Technology (NIST), Gaithersburg, MD 20899, USA}
\author{Veronica G. Reynolds}
\affiliation[ucsb]{Materials Department, University of California, Santa Barbara, CA 93106, USA.}
\author{Subhrangsu Mukherjee}
\affiliation[nist]{Material Measurement Laboratory, National Institute of Standards and Technology (NIST), Gaithersburg, MD 20899, USA}
\author{Daniel Sunday}
\affiliation[nist]{Material Measurement Laboratory, National Institute of Standards and Technology (NIST), Gaithersburg, MD 20899, USA}
\author{Tyler B. Martin}
\affiliation[nist]{Material Measurement Laboratory, National Institute of Standards and Technology (NIST), Gaithersburg, MD 20899, USA}
\author{Peter A. Beaucage}
\affiliation[nist]{NIST Center for Neutron Research, National Institute of Standards and Technology (NIST), Gaithersburg, MD 20899, USA}
\author{Michael L. Chabinyc}
\affiliation[ucsb]{Materials Department, University of California, Santa Barbara, CA 93106, USA.}
\author{Dean M. DeLongchamp}
\affiliation[nist]{Material Measurement Laboratory, National Institute of Standards and Technology (NIST), Gaithersburg, MD 20899, USA}
\email{dean.delongchamp@nist.gov}
\author{Adarsh Krishnamurthy}
\affiliation{Department of Mechanical Engineering, Iowa State University, Ames, IA 50010, USA}
\author{\\Baskar Ganapathysubramanian} 
\email{baskarg@iastate.edu}
\affiliation{Department of Mechanical Engineering, Iowa State University, Ames, IA 50010, USA}

\begin{document}
\begin{abstract}
    Polarized Resonant Soft X-ray scattering (\prsoxs) has emerged as a powerful synchrotron-based tool that  combines principles of X-ray scattering and X-ray spectroscopy. \prsoxs~provides unique sensitivity to molecular orientation and chemical heterogeneity in soft materials such as polymers and biomaterials. Quantitative extraction of orientation information from \prsoxs{} pattern data is challenging because the scattering processes originate from sample properties that must be represented as energy-dependent three-dimensional tensors with heterogeneities at nanometer to sub-nanometer length scales. We overcome this challenge by developing an open-source virtual instrument that uses Graphical Processing Units (GPUs) to simulate \prsoxs{} patterns from real-space material representations with nanoscale resolution. Our computational framework --~called CyRSoXS (\url{https://github.com/usnistgov/cyrsoxs})~-- is designed to maximize GPU performance, including algorithms that minimize both communication and memory footprints. We demonstrate the accuracy and robustness of our approach by validating against an extensive set of test cases, which include both analytical solutions and numerical comparisons, demonstrating a speedup of over three orders relative to the current state-of-the-art \prsoxs{} simulation software. Such fast simulations open up a variety of applications that were previously computationally infeasible, including (a) pattern fitting, (b) co-simulation with the physical instrument for \textit{operando} analytics, data exploration, and decision support, (c) data creation and integration into machine learning workflows, and (d) utilization in multi-modal data assimilation approaches. Finally, we abstract away the complexity of the computational framework from the end-user by exposing CyRSoXS to Python using Pybind. This eliminates input/output (I/O) requirements for large-scale parameter exploration and inverse design, and democratizes usage by enabling seamless integration with a Python ecosystem (\url{https://github.com/usnistgov/nrss}) that can include parametric morphology generation, simulation result reduction, comparison to experiment, and data fitting approaches.

% Need for a companion virtual simulator. \\ Key claims: 
% \begin{itemize}
%     \item Accomplish near real time simulation of RSoXS
%     \item Use GPU acceleration to get 1000X over current state-of-art. Careful memory management and communication minimization important aspects. 
%     \item Additional functionality: multi-material, python binding, open-source with future road-map of features, 
%     \item Extensive set of validation examples.
%     \item Tutorials and python binding to deliberate democratization of virtual instrument tools.
% \end{itemize}
\end{abstract}
% \end{frontmatter}
\section{Introduction}
%\section{Introduction}
\renewcommand{\thefootnote}{\fnsymbol{footnote}}
Developing process-structure-property relationships is a central pillar of material science and engineering research. Understanding the effect of composition, structure, and processing on the performance of a material can enable the intelligent and efficient tuning of the process variables to improve the end performance of the material in a given application. With these process-structure-property relationships, the exciting goal of designing new materials instead of discovering them becomes a reality. Thus, there is an ever-present need to develop new characterization methods to elucidate material structure with increasing detail and clarity.

Structural characterization is particularly challenging in soft matter due to its semi-disordered nature. Some important aspects of soft material structure include spatial heterogeneities in composition, density, molecular orientation/conformation, and the degree of order. Recent advancements in synthesis and materials processing have unlocked access to systems in which all aspects of soft material structure might ultimately be controlled by design. However, despite an enormous acceleration in the capability and speed of characterization methods across many length scales, it remains a fundamental and pervasive challenge to efficiently, rigorously, and robustly assimilate materials structure characterization data streams into a self-consistent \textit{digital twin} that describes material structure. If achieved, the resultant comprehensive structural description would allow us to understand, predict, and eventually control how material properties arise from a complex interplay of different aspects of structure across relevant length scales.
%beltran2019computational,wessels2021machine

In this context, there have been recent efforts to integrate computational tools with experimental data streams. \textit{Virtual instruments} that mimic the physical principles of the characterization method---X-ray diffraction, light spectroscopy, electron transmission~\citep{wessels2021computational,mukherjee2021polarized,pryor2017streaming,reynolds2022simulation}---can transform how downstream analysis of experimental data streams is performed. For instance, a virtual instrument can enable rapid data quality evaluation and provide statistically rigorous estimates of when enough data has been collected. Such approaches can maximize the utilization of heavily-subscribed instruments at centralized facilities such as X-ray and neutron sources. Furthermore, a virtual instrument can allow principled down-selection of plausible hypotheses for developing structure-property relationships. Such virtual tools also allow formal analysis and characterization of uncertainty, identify the most sensitive features, and allow \textit{in silico} experimentation before performing physical experiments for greater efficiency in experiment execution. Finally, the success of artificial intelligence and machine learning (AI/ML) methods~\citep{axelrod2022learning,vasudevan2021machine,guo2021artificial,gomes2019artificial} point to the possibility of integrating experimental data with the virtual instrument to provide automated and formal approaches to assimilating complementary data streams---for instance, real space (electron microscopy) and frequency space (X-ray diffraction)---to create a self-consistent and multimodal digital twin. 

Polarized Resonant Soft X-ray scattering (\prsoxs) is a recently developed technique with unique characterization abilities \citep{collins2022resonant} and an excellent candidate for developing a virtual instrument. Typical scattering experiments performed at hard X-ray energies provide a very low contrast between organic constituents in a material. \prsoxs~ overcomes this limitation by combining conventional small-angle X-ray scattering (SAXS) with soft X-ray spectroscopy to yield a tunable scattering contrast. The energies of this soft X-ray are scanned across absorption edges of the light elements (C, N, O), commonly found in organic materials, often yielding significant contrast variation and substantially improved signal-to-noise ratio for organic systems. \prsoxs~ thus provides a path to probe the structure in the $\mathrm{nm - \mu m}$ range with both chemical and physical sensitivity without the need to perturb the system with labels such as the heavy element "stains" commonly used to enhance SAXS or the radioisotopes commonly used to enhance small angle neutron scattering (SANS). The contrast enhancement makes P-RSoXS particularly useful for probing the structure of thin ($<$~200~nm) films, samples that are challenging for hard X-rays and neutrons due to the small scattering volumes. Composition contrast with P-RSoXS is so significant that short exposures of thin films -- less than 1 min at normal incidence -- at resonant energies with high contrast will yield patterns with quality similar to conventional bulk SANS patterns requiring mm-scale  sample volumes and hours to collect. The approach also does not require grazing-incidence geometries which are commonly used to gain signal in the X-ray scattering of thin films.

The variable sensitivity of \prsoxs~to each chemical bond can amplify scattering intensity even with only small chemical differences between materials, which enables the extraction of useful structure information for heterogeneous materials. A unique aspect of \prsoxs~ is that it is sensitive to molecular orientation via interaction of the soft X-ray electric field vector with oriented transition dipoles within the sample. Complex \prsoxs~ patterns can arise from orientational heterogeneities. This unique aspect of \prsoxs~ provides exciting opportunities for characterizing previously unmeasurable aspects of the structure of soft materials, but it makes adapting conventional SAXS or SANS analysis approaches nearly impossible because the materials properties that affect contrast in those techniques are effectively scalar quantities. A new analysis framework is required to represent independent fluctuations in material composition and molecular orientation on sub-nanometer length scales. The availability of a virtual analog to \prsoxs~ will enable the discovery and quantification of structure in complex, chemically heterogeneous soft systems. Motivated by this exciting promise, here we will describe our development of CyRSoXS -- a fast, graphics processing unit (GPU) accelerated virtual instrument for  \textit{\textbf{Polarized Resonant Soft X-ray scattering}} (\prsoxs). 

To dispel any question regarding which technique we address herein, we note that, because \prsoxs~ is not yet a mainstream technique, a variety of different acronyms have been proposed for it, including "R-SoXS" and "PAXS."\citep{gann2016origins} The community now appears to have settled on "RSoXS" and "\prsoxs." It is not uncommon for practitioners to use only "RSoXS" when exploiting its composition contrast capabilities and to use "\prsoxs" when adding its orientation contrast capabilities. We should mention, however, that these contrast modes are intrinsically linked. It is not possible to perform RSoXS without polarization and its concomitant molecular orientation sensitivity. Even circular polarization will yield patterns that can be significantly affected by molecular orientation effects. These principles suggest that model-free, composition-only analyses of \prsoxs~in systems having significant but ignored molecular orientation fluctuations may yield incorrect results, a situation that could be greatly improved with a fast virtual instrument.

\noindent\textbf{Current state-of-art:} 
The current state-of-the-art \prsoxs{} simulator, developed by \citet{gann2016origins} in Igor Pro\footnote{Certain commercial equipment, instruments, or materials are identified in this paper in order to specify the experimental procedure adequately.  Such identification is not intended to imply recommendation or endorsement by NIST, nor is it intended to imply that the materials or equipment identified are necessarily the best available for the purpose.} has been pivotal in answering many scientific questions~\citep{jiao2017quantitative,ye2016manipulation,song2018highly,song2019efficient,mukherjee2017morphological,litofsky2019polarized}. However, it has limitations on practical deployment in terms of speed and no opportunity for deployment on the state-of-the-art high performance computing (HPC) clusters. These limitations become most apparent when attempting to fit experimental data using goal-seeking algorithms that adjust material structure input parameters to obtain agreement between simulation and experiment. Such optimization routines require a significant number of forward simulations and thereby motivate the need for the fast forward simulator. The commercial licensing of Igor Pro further hinders the democratization and availability of the tool to many researchers. There has been rapid growth in the interest in leveraging advances in Machine Learning and Data Analytics among material scientists for material design and exploration. The availability of a fast forward simulator is a critical necessity for data creation and integration into Machine Learning Model Operationalization (MLOps) workflows. More practically, since Python has become the de-facto language for data analysis and Machine Learning, the currently available simulator does not provide any straightforward integration for researchers to utilize such tools.

\noindent\textbf{Our contributions:} We build upon an earlier framework that modeled the physics of soft X-ray scattering through a heterogeneous thin film~\citep{gann2016origins}. In particular, we significantly speed up execution time and, via integration with a Python ecosystem, incorporate substantial additional functionality. Our key contributions in this paper are:
\begin{itemize}
    \itemsep0em 
    \item Accomplish near real-time simulation of RSoXS at sizes/resolutions (up to $2^{28}$ or 268 million voxels\footnote{On \textsc{V100} GPUs. This size is limited purely by the GPU memory (\secref{sec: Algorithm})}) that were hitherto not possible. 
    \item Use GPU acceleration to achieve 1000 $\times$ speedup over current state-of-art approaches. This is achieved by careful design of ``GPU-friendly'' data structure and algorithms--- including memory and communication considerations.
    \item Careful software design of a simulation engine that lies at the center of  a feature-rich data analysis and model exploration ecosystem when combined with Python code bases for morphology generation, simulation result reduction, and data-fitting. Python binding democratizes access, simplifies usage, and enables seamless integration with AI / ML libraries and eliminates the bottleneck of I/O operation\footnote{This can quickly become a bottleneck as File I/O is several orders of magnitude slower than memory read/write.}, especially during parameter exploration or inverse design.
    \item A new and well-documented voxel-based material structure data file format in Hierarchical Data Format-5 (HDF5) that includes capabilities for verbose metadata, multiple materials, independent representation of composition and orientation, and an intuitive Euler angle description of material orientation. 
    \item An extensive set of validation examples developed by a growing community across multiple institutions.
    \item Tutorials that serve as unit tests for this open-source framework. The full software stack is open-source and requires access to CUDA-enabled hardware. %This contrasts the current state-of-the-art framework, which requires Igor commercial licence.}
\end{itemize}

The rest of the paper is organized as follows: We begin by briefly introducing \prsoxs{} in \secref{sec: prsoxs}, followed by a detailed mathematical model in \secref{sec:Math}; We detail the data structures and algorithms in \secref{sec: Algorithm}, and present results including validation cases in \secref{sec:results}. We show the performance of CyRSoXS with varying problem size and scaling to multiple GPUs  in \secref{sec:Performance}. We discuss integration with Python environments in \secref{sec:pybind}, and we conclude by discussing the implications and path for future developments in \secref{sec:Conclusion}.
%In this work, we present a Graphical Processing Unit (GPU) accelerated virtual instrument for \prsoxs~ measurement. GPU has been traditionally been used for accelerating the computer graphics. With the advent of GPGPU (General purpose Graphical Processing Unit) programming~\citep{nvidia2011nvidia,munshi2011opencl}, GPU has been successfully deployed for solving the problems across various field ranging from the atomistic scale~\citep{das2019fast,ziogas2019data} simulations to the astronomical scale~\citep{fernando2018massively,potter2017pkdgrav3}. Recent development in GPU technologies have enabled to accelerated codes by several order of magnitudes.  These advances have driven the development of softwares capable of efficiently using and unifying the disparate architectures. Further, the readily availability of GPU devices, both on a big compute cluster (\Summit~, \Frontera) and workstations has made GPU a viable choice. The availability of these fast instrumentation, can democratize  pathways  for accelerated  materials  discovery  via  computational  material science.

\section{Polarized Resonant Soft X-ray Scattering (\prsoxs)} \label{sec: prsoxs}
\begin{figure}[b!]
    \centering
    \includegraphics[width=\linewidth]{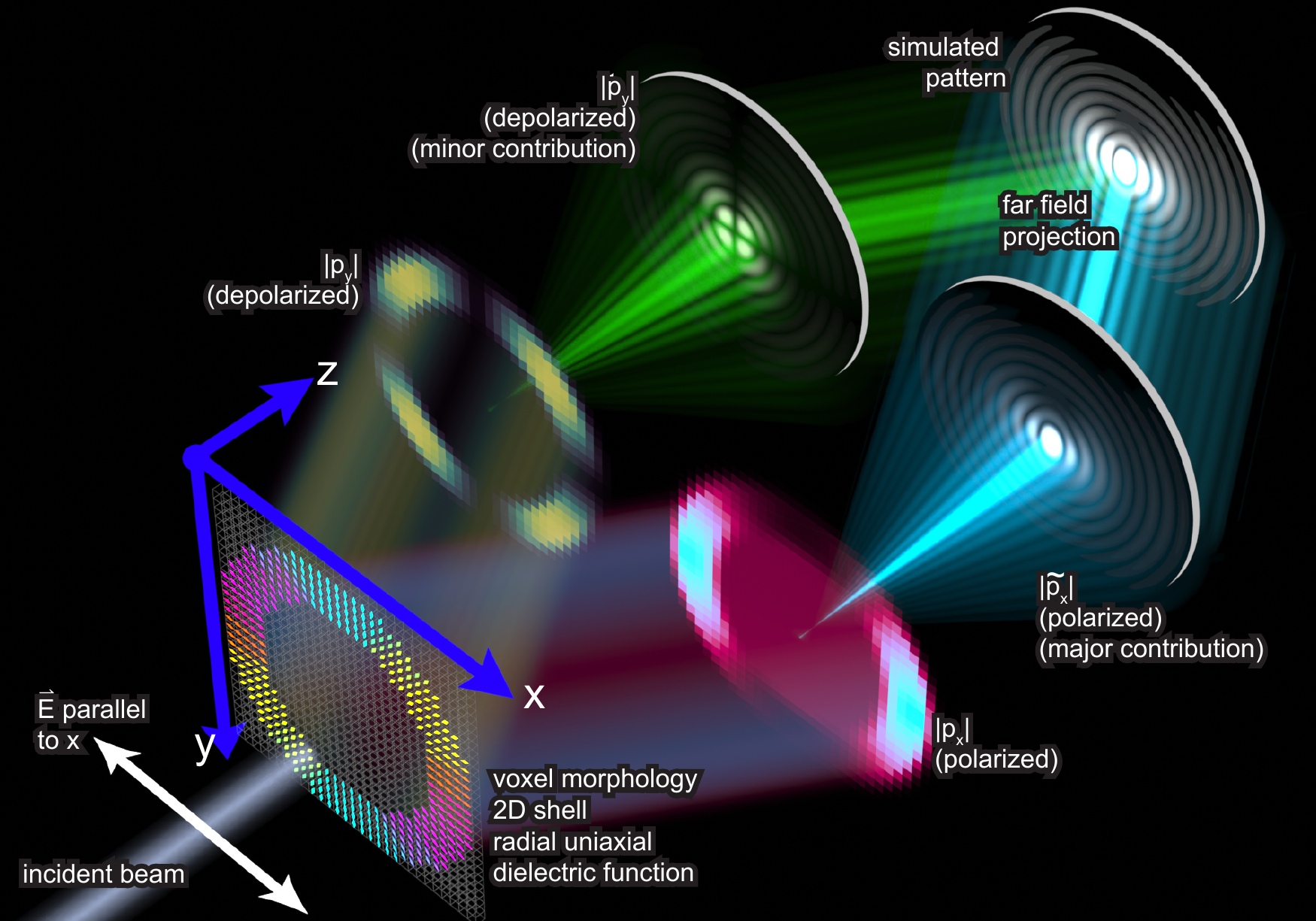}
    \caption{Schematic of \prsoxs, where a polarized soft X-ray beam passes through a sample, interacting with and scattering off the electrons in that sample; these scattered X-rays are collected on an X-ray sensitive detector.}
    \label{fig: schematic}
\end{figure}

% The following paragraph might require some additional explanations to be a bit more clear

In \prsoxs, a polarized soft X-ray beam passes through a sample, interacting with and scattering off the electrons in that sample; these scattered X-rays are collected on an X-ray sensitive detector (typically charge coupled device (CCD) or complementary metal oxide semiconductor (CMOS)). \figref{fig: schematic} shows a condensed version of the physical principles of \prsoxs, which are explained in greater detail in a recent comprehensive review of the technique and its application to soft materials.\citep{collins2022resonant} This photon-electron interaction strength depends on the X-ray energy and the chemistry of the molecules within the sample. At energies far from an absorption edge, X-rays interact equally with all electrons in the sample, and the interaction strength scales directly with the electron density. Near an absorption edge, the interaction strength increases dramatically when the incident X-ray energy is commensurate with the energy required to resonantly excite an electron to an unoccupied molecular orbital. The K-absorption edge of many lightweight elements (C, O, N, F) lies in the soft X-ray energy regime ($100$~$eV\lesssim E_{photon} \lesssim ~2$~$keV$); all are commonly exploited in \prsoxs. Selection of the incident energy near the core binding energy of the electrons makes the technique element-specific, whereas the chemical bonds that define the excited state unoccupied molecular orbital energy make the technique sensitive to specific bonds or moieties. The spectroscopic scattering pattern thus provides a tunable, chemically-sensitive probe of nanoscale and mesoscale components in a heterogeneous complex material~\cite{attwood2017x}.

The resonant soft X-ray absorption is described by a transition dipole moment that couples the initial and final states. The initial state of the electron is a core orbital that is spherically symmetric, therefore the geometric dependence of the interaction strength is defined by the unoccupied molecular orbital, which for most soft X-ray resonances can be represented as vectors or planes parallel or perpendicular to the bond \cite{stohr1992nexafs}. Soft X-ray absorption, a principal contributor to scattering contrast, varies as the dot product of the electric field vector and the transition dipole moment. This interaction makes \prsoxs~ sensitive to spatial distributions in molecular orientation. For instance, in the case of carbon fused ring compounds, when the X-ray energy is in resonance with the fundamental carbon electron transition (C1$s\xrightarrow{}\pi^*$), the molecules exhibit vector transition dipole moments perpendicular to the ring planes \cite{mannsfeld2012tune}. Two identical molecules oriented differently within a sample will have different interaction strengths with a fixed electric field vector, and there will be a scattering contrast between them. If the orientation of these molecules is spatially correlated in a sample, a scattering pattern will be observed. For example, \prsoxs{} can detect correlated interfacial molecular orientation regions (such as mixtures of amorphous, semi-crystalline, or liquid crystalline phases). Crystalline, semi-crystalline, and liquid-crystalline organic materials have locally large anisotropic bond orientation statistics, impacting the mechanical, optical, and electronic properties of these materials. Understanding these relative orientations at different length scales is necessary for a detailed understanding of organic thin-film devices. In addition, \prsoxs{} had been shown to reveal local molecular alignment independent of overall crystallinity and represents an essential new tool for understanding the structure-property relationship and examining the connection between transport properties and morphology in organic and hybrid organic-inorganic electronic devices.

The interaction of X-rays with a system can be encoded using a 3D analog of the complex index of refraction, $\Tensor{N}$. Each component of this tensor is a function of the dispersive and absorptive components of the index of refraction, $ N_{ij}(E)= f(\delta(E), \beta(E))$, where E is the photon energy, $\delta$ is the dispersive component, and $\beta$ is the absorptive component of the index of refraction. In the hard X-ray regime, far away from the resonance frequency of the constituent atoms, the real part of the complex index of refraction is a scalar proportional to the electron density of the material. The imaginary part is negligible due to low absorption, and the electron density difference between constituent materials determines the scattering contrast of the system. However, close to the absorption edge of the constituent atoms, the electrons get excited to the unoccupied molecular states or vacuum, and therefore $\beta$ will naturally exhibit peaks and other absorption features that will differ depending on orientation; changes in $\delta$ are also expected due to causality and can be calculated using the Kramers-Kronig relations~\cite{wang2010resonant,watts2014calculation}.

\section{Mathematical Model}
\label{sec:Math}
\subsection*{Notation}
\begin{tikzpicture}
\node at (0, 0) {\framebox{
    {\begin{varwidth}{0.95\linewidth}
    \begin{itemize}[itemsep=0pt]
    \item Vectors are represented as lower case bold letters with a single underline, $\Vector{e},~ \Vector{p}$
    \item Tensors (or specifically, matrices) are represented as uppercase bold letters with double underline, $\Tensor{R},~\Tensor{N}$
    \item Scalars are represented as lower case letters, $\phi,~~n_x$
    \item Counting (over components) integer is $j$
    \end{itemize}
    \end{varwidth}}
}};
\end{tikzpicture}

Having described the overall mechanism of \prsoxs, we next detail the mathematics of the simulation. 
% \subsection{Notations}

% \subsection{Inputs}
% The overall simulation is defined by global and local input variables. Input1 shows the global variables. The input morphology is defined in terms of voxels, where each voxel represents the corresponding component volume fraction along with the biaxial director vector field.  

% % 

% 
%%%%%%%%%%%%%%%%%%%%% Maths %%%%%%%%%%%%%%%%%%%%%%%
\subsection{Morphology:}
Consider a morphology composed of a $c$ component mixture. We discretize the morphology into a uniformly spaced voxel grid. Each voxel contains some (or all) of the $c$ components. Each of these components can either be amorphous or can be oriented. If a component is oriented, we assume that it is well represented by a uniaxial representation.~\footnote{While the uniaxial representation is adequate for most use cases currently considered, it has disadvantages. A key disadvantage is that it will convey less information than other representations. It cannot perfectly represent properties at the molecular level: the simplest representations of molecular-level properties for most molecules would be biaxial. It cannot represent complex orientation distributions at the sub-voxel level: only a single orientation mode is conveyed per material, and the "shape" of the distribution is lost.}

A key advantage of the uniaxial assumption is that it is simple and allows the construction of a simple, abstract model of properties within a voxel. This abstract model and associated data structures are independent of the material and energy. The abstract model can then be combined with a material library, which can be stored in memory, to allow the same model to be re-used for different materials and different energies. This abstract representation requires only two scalar fractions (volume fraction and orientation fraction) and two Euler angles per material/voxel for the uniaxial assumption. In contrast, a biaxial representation would require five scalar fractions (volume fraction and four orientation mixing parameters) with three Euler angles for a similar abstract model. To represent arbitrary distributions of a biaxial representation in an abstract model would require including non-diagonal elements of the full tensor for 19 unique scalar fractions (volume fraction, six elements with 3 coefficients each on the original "molecular" biaxial elements), significantly increasing memory and communication footprints. %and making the structure almost completely opaque to visual examination and cursory error checking.

A uniaxial representation conveys the necessary properties for materials with a single dominant orientation mode of one particular molecular axis, which we judge to be a large number of use cases. If a more faithful representation of a multimodal orientation distribution is required, that can be approached in our framework using by breaking a component into additional materials with identical dielectric functions and volume fractions that add to the total for that component, but which have distinct orientations reflecting the expected distribution. If a more faithful representation of molecular-level properties is required, it is possible to use a system of uniaxial functions to represent an underlying biaxial function, but that is not a currently supported use case for our approach. We will lay out a clear pathway to relax this assumption and consider biaxial representation in the next release version of CyRSoXS. %Our future software development roadmap relaxes this assumption to incorporate biaxial representation. 

Each voxel, therefore, has four features associated with each component $j = 1 \dots c$:
\begin{itemize}
    \item $v_{frac}^j$: fraction of volume occupied by component $j$ in this voxel. By definition, 0 $\leq$ $v_{frac}^j$ $\leq$ 1, and the sum $v_{frac}^j$ across all $j$ (that is, the sum of all volume fractions of all materials) within a voxel is expected to be 1.0, CyRSoXS will not check whether they sum to one, although such checking can be done with model visualization code provided in our broader Python ecosystem. We encourage as a best practice the use of vacuum as an explicit material in the model, such that model self-consistency is straightforward to confirm.  
    \item $s^j$: degree of alignment of component $j$ in this voxel. This parameter indicates the volume fraction of component $j$ that is oriented (as opposed to unaligned). We expect that 0 $\leq$ $s^j$ $\leq$ 1, but unlike $v_{frac}^j$, there is no expectation of any constraint involving other materials. $s^j$ is a relative volume fraction; in other words $s^j$ is multiplied by $v_{frac}^j$ to yield the absolute volume fraction of oriented material j in a voxel (see Eq.~\ref{eq: ncorr} below). This parameter is conceptually identical to the well-known uniaxial "orientational order parameter," $S$, but only in the range of 0~$\leq$~$S$~$\leq$ 1, where $S$ = 1 indicates complete alignment with a director (our director is defined by the Euler angles described below) and $S$ = 0 indicates an isotropic condition. We note that the orientational order parameter $S$ can also include the range -0.5~$\leq$~$S$~$\leq$~0, which indicates orientation perpendicular to the director, but we do not support values of $s^j$ less than zero. Expressing perpendicular orientations should instead be accomplished by explicit adjustment of Euler angles.
    \item $\phi^j$: orientation feature 1, defined as the (first) rotation of component $j$ about the Z-axis.
    \item $\theta^j$: orientation feature 2, defined as (second) rotation of component $j$ about the (original) Y-axis.
\end{itemize}
The last two features represent the Euler angle representation of material orientation in a voxel. 

We refrain from providing overly prescriptive guidance on model design because there may be use cases for CyRSoXS that we cannot anticipate. However, we offer here some model design choices that have worked well for our internal testing and for many of our validation cases provided in \secref{sec:results}. Most models will represent the real-space structure of a thin film, so they will typically have larger $x$ and $y$ "lateral" dimensions and a smaller $z$ "height" dimension. We consider it a best practice for $x$ and $y$ to have the same dimensions and resolutions. Common $x$ and $y$ dimensions (meaning the length of the whole model on a lateral side) are micron scale, perhaps ranging from (0.5~to~5)~$\mu$m. Common lateral resolutions include 512~$\times$~512, 1024~$\times$~1024, and 2048~$\times$~2048, although larger sizes are possible. The $z$ resolution is usually a smaller multiple of 2; common values include 32, 64, and 128. The voxels are considered perfect cubes in CyRSoXS, such that the model dimensions are the product of the resolution and the length of a voxel side. These model dimensions and resolutions correspond to voxels with side lengths in the (0.2~to~10)~nm range.

These model dimensions and resolutions are compatible with data fusion workflows where real-space images derived from atomic force microscopy (AFM), transmission electron microscopy (TEM), or other imaging methods are used as a foundation for CyRSoXS model creation. In some cases such images could be used to assign $v_{frac}^j$ across voxels for different components, depending on the contrast mode of the imaging. The other voxel-level parameters, $s^j$, $\phi^j$, and $\theta^j$ will most likely not be available from imaging methods because there is a lack of techniques that are sensitive to molecular orientation in soft materials at the nanoscale. (This fact provides much of our motivation for investment in \prsoxs~ interpretation!) Hypothesis driven parametric assignment of $s^j$, $\phi^j$, and $\theta^j$ might instead be employed. Models built entirely parametrically are certainly possible, as demonstrated for many of our validation cases shown in \secref{sec:results}. 

\begin{figure*}[t!]
    \begin{subfigure}{0.2\linewidth}
        \centering
        \includegraphics[width=\linewidth]{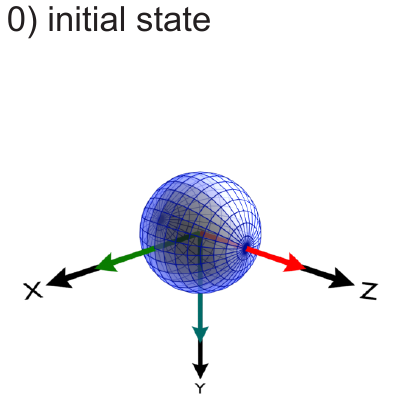}
        \subcaption{Initial state}
        \label{fig:Euler0}
    \end{subfigure}
    \begin{subfigure}{0.2\linewidth}
        \centering
        \includegraphics[width=\linewidth]{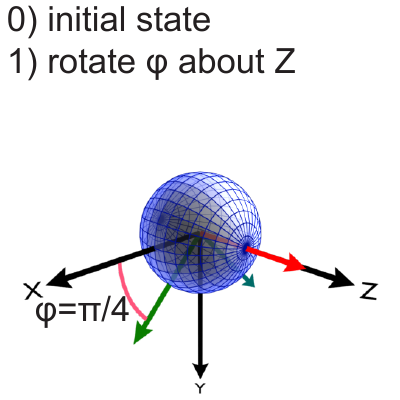}
        \subcaption{$\phi$ around Z}
        \label{fig:Euler1}
    \end{subfigure}
    \begin{subfigure}{0.2\linewidth}
        \centering
        \includegraphics[width=\linewidth]{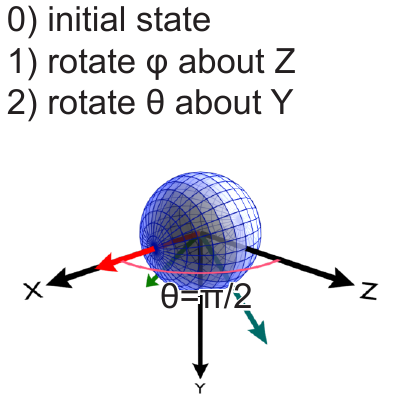}
        \subcaption{$\theta$ around Y}
        \label{fig:Euler2}
    \end{subfigure}
    \begin{subfigure}{0.2\linewidth}
        \centering
        \includegraphics[width=\linewidth]{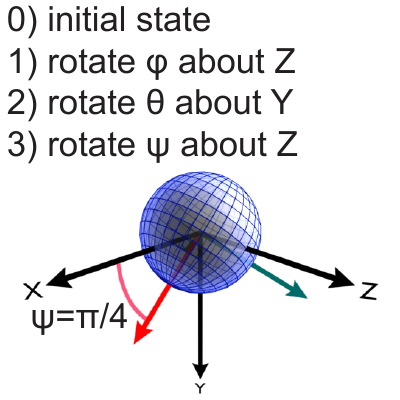}
        \subcaption{$\psi$ around Z}
        \label{fig:Euler3}
    \end{subfigure}
     \caption{Different steps of the Euler angle rotation.}
     \label{fig:EulerRotation}
 \end{figure*}

\noindent \textbf{A brief primer on Euler angles:} For Euler angles, we use the ZYZ convention. We assume that the primary alignment axis starts parallel to Z - axis (0,0,1) (\figref{fig:Euler0}). This is also the default direction of the simulated incident beam. According to this convention (\figref{fig:EulerRotation}), and with reference to the rotation matrices B, C, and D, which are further defined in Equation \eqref{rotBCD}:
\begin{enumerate}
    \item The first rotation is by an angle $\varphi$ about the Z-axis using rotation matrix D. (\figref{fig:Euler1}) %Note that because we consider a uniaxial dielectric function, rotation about $\varphi$ does not change the alignment vector. Therefore, it is ignored and is assumed to be 0.
    \item The second rotation is by an angle $\theta$ about the original Y-axis using rotation matrix C.(\figref{fig:Euler2})
    \item The third rotation is by an angle $\psi$ about the original Z-axis using rotation matrix B.(\figref{fig:Euler3})
\end{enumerate}
%  \begin{figure}[h]
%      \centering
%      \includegraphics[trim=0 0 0 10, clip,scale=0.5]{Figures/rotation.png}
%      \caption{Euler angle rotation. Figure taken from \cite{weisstein2009euler}}
%      \label{fig:EulerRotation}
%  \end{figure}
%  The rotation matrices are given as:
% \begin{equation}
% \begin{split}
%  D = & \begin{bmatrix}
% \cos{\phi} & \sin{\phi} & 0\\
% -\sin{\phi} & \cos{\phi} & 0 \\
% 0 & 0 & 1 \\
% \end{bmatrix}   \\
%  C = & \begin{bmatrix}
% 1 & 0 & 0\\
% 0 & \cos{\theta} & \sin{\theta}\\
% 0 & -\sin{\theta} & \cos{\theta}
% \end{bmatrix}   \\
% B = & \begin{bmatrix}
% 1 & 0 & 0\\
% 0 & 1 & 0\\
% 0 & 0 & 1
% \end{bmatrix}   
% \end{split}
% \label{eq:rotMatrix}
% \end{equation}

We note that other conventions are possible and have been used in the literature; for instance, ~\citet{gann2016origins} used the vector orientation in 3D space to define an equivalent morphology. These equivalent conventions can be easily transformed into the Euler angles using suitable rotation transformations. A benefit of our convention is its straightforward expandability into a biaxial representation by adding a third Euler angle.

% In order to completely define a biaxial system, an additional degree of alignment $u^j,p^j,f^j$, commonly known as SUPF ordering~\cite{rosso2007orientational} is needed. However, in this work, we consider biaxial system as a two component uniaxial system, which is discussed in ~\secref{sec: Biaxial}.

% A biaxial director vector is associated with each component, given by $\Vector{s}_j^1$ and  $\Vector{s}_j^2$, along with the corresponding volume fraction filled with isotropic material $\phi_{j}^u$. The biaxial director vector ($\Vector{s}_j^1$ and  $\Vector{s}_j^2$) determines the fraction of voxel volume which is taken by the aligned fraction of the material, whereas $\phi_{j}^u$ determines the fraction of voxel volume that is isotropic. The overall volume fraction for a given material can be given by:
% \begin{equation}
%     \phi_j = \phi^u_j + |\Vector{s_j}^1| + |\Vector{s_j}^2|
% \end{equation}

% \begin{remark}
% A uniaxial systems can be defined fully, only by specifying $\Vector{s}_j^1$.
% \end{remark}

\subsection{Material properties:}
As mentioned in \secref{sec: prsoxs}, the interaction of soft X-rays with a material is encoded in the 3D analog, $\Tensor{N}$, of the material-specific, complex index of refraction. $\Tensor{N}$ is a $ 3 \times 3$ data structure that exhibits energy dependence. For a uniaxial system, $\Tensor{N}$ can be diagonalized as:
\begin{equation}
\Tensor{N} = 
\begin{pmatrix}
n_{\parallel}& 0 &  0 \\
        0 &n_{\bot}& 0 \\
        0 & 0 &n_{\bot}
\end{pmatrix}
\label{eq: uniaxialRefrativeIndex}
\end{equation}
We will refer to $\Tensor{N}$ as the refractive index for brevity, with the understanding that it actually is a convenient 3D analog to the complex index of refraction.

\subsection{Mathematical representation of \prsoxs}\label{sec: MathSteps}
The mathematical operations that mimic \prsoxs{} can be divided into 6 steps:
\begin{enumerate}
    \item \textit{\textbf{Effective Refractive Index:}} For each voxel, the effective refraction tensor for material component $j$ can be computed using the aligned and unaligned fraction as:
    \begin{equation}
    \footnotesize
    \Tensor{N_{eff}}^{j} = v^{j}_{frac} \bigg(\underbrace{s^j\Tensor{N}^j}_{\text{aligned part}} + \underbrace{(1 - s^{j})  \frac{1}{3} {\text{Trace}}(\Tensor{N}^j) \Tensor{I}}_{\text{unaligned part}}\bigg)
    \label{eq: ncorr}
    \end{equation}{}
    % This step is independent of either uniaxial or biaxial assumptions. 

\item \textit{\textbf{Rotated Refractive Index:}} For each material component, $j$, in every voxel, the effective refractive tensor, $\Tensor{N_{eff}}^{j}$, is rotated according to the alignment vector, $\Tensor{R}^j$:
\begin{equation} \label{rotBCD}
\Tensor{R}^j = \Tensor{B}^j \quad \Tensor{C}^j \quad \Tensor{D}^j
\end{equation}
where $\Tensor{B}$, $\Tensor{C}$ and $\Tensor{D}$ are the rotation matrices following the Euler angle convention depicted in ~\figref{fig:EulerRotation}.
The rotated refractive index, $\Tensor{N_{rot}}^{j}$ is computed as:
    \begin{equation}
        \Tensor{N_{rot}}^{j} = \Tensor{R}^{j^T} \quad  \Tensor{N_{eff}}^{j} \quad \Tensor{R}^{j}
        \label{eq: nrot}
    \end{equation}

 \item \textit{\textbf{Polarization Computation:}} The induced molecular polarization $\Vector{p}$ produced by the electric field $\Vector{e}$ of the beam is computed as:
 \begin{equation}
   \Vector{p} = \sum_{j = 1}^c\frac{(\Tensor{N_{rot}}^j:\Tensor{N_{rot}}^j - \Tensor{I})\cdot\Vector{e}}{4\pi}  
   \label{eq: polarization}
 \end{equation}
 %  \begin{equation}
%   \Vector{p}(\Vector{x_p}) = \sum_{j = 1}^c\frac{(\Tensor{N_{rot}}^j:\Tensor{N_{rot}}^j - \Tensor{I})\cdot\Vector{e}}{4\pi}  
%   \label{eq: polarization}
%  \end{equation}
Differences in the $\Vector{p}$ components are the origin of scattering contrast in \prsoxs. The structure of these components in real space can be complex, even for simple structures. For a qualitative picture of this complexity, \figref{fig: schematic}~ shows an illustration of $p_x$ and $p_y$ magnitudes for a simple, compositionally homogeneous disk with radial orientation of a uniaxial dielectic function (polyethylene in this image), at an energy that enhances orientation contrast in the material.

We include a switch to allow the the final scattering pattern to be computed by averaging across different orientations of the electric field. If this computation is enabled, it  is performed as follows: we start with $\Vector{e} = (1, 0, 0)$, we rotate $\Vector{e}$ using a rotation matrix $\Tensor{U}$. We compute the polarization using Eq.~\ref{eq: polarization}, which is then rotated back using $\Tensor{U}^T$ (i.e $\Vector{p} \leftarrow \Tensor{U}^T\cdot\Vector{p}$). The rotation is done in fine increments across a range, and then averaged. This rotation functionality is included as a capability to smooth simulated pattern features that are due to the finite size of the models. Rotating in small increments and averaging the scattering pattern in this way effectively simulates a noninteracting polydomain material where each domain is a copy of the original model that is rotated about the z axis. Enabling this functionality will better capture electric field interactions with model details. This functionality should not be used, however, if a model has structural features that are intentionally nonuniform in x-y plane directions.

%To perform averaging, one can either rotate the material representation, or equivalently rotate the e-field. Our computations are performed by rotating \Vector{e} (see section XX). Considering a rotation matrix, $\Tensor{U}$ of the electric field, the $\Vector{p}$ for this rotation is calculated as $\Tensor{U}\cdot \vector{e}$  %Under rotation, $\Vector{e}$ is first rotated back to the reference orientation from its current orientation, the induced polarization is then computed. The induced polarization is then rotated back to the original orientation for averaging. 
%After $\Vector{p}$ is calculated, it is rotated into the frame of reference of $\Vector{e}$, such that 

%  where $\Vector{x_p}$ denotes the frame of reference that is rotated along with the rotation of $\Vector{e}$ such that $\Vector{e}$ is always parallel to $\Vector{x}$ and $\Vector{k}$ is always parallel to $\Vector{z}$.
 \item \textit{\textbf{Fast Fourier Transform (FFT):}} To get the reciprocal space ($\Vector{q}$) representation, we first compute the FFT of the real space polarization vector $\Vector{p}$:
 \begin{equation}
     \Tilde{\Vector{p}} = \texttt{FFT}(\Vector{p})
 \end{equation}

 \item \textit{\textbf{Scatter computation:}} 
 The differential scattering cross-section, $X(\Vector{q})$ is given by:
 \begin{equation}
     X(\Vector{q})= \vert|\Vector{k}^\text{\textbf{out}}|^2(\Tensor{I} - \UnitVector{r}\UnitVector{r})\cdot\Tilde{\Vector{p}}\vert^2
      \label{eq: 3Dscatter}
 \end{equation}
 where $\UnitVector{r}$ is the real-space unit vector from the sample to the detector, such that $\Vector{r} \approx \Vector{k}^\text{\textbf{out}} = \Vector{k}^{\text{\textbf{in}}} + {\Vector{q}}$, and $\Vector{k}^{\text{\textbf{in}}}$ is the wavenumber of the incident wave. \eqnref{eq: 3Dscatter} is derived using the first order Born approximation (far-field limit)~\cite{born2013principles}.

 The individual components of $\Tilde{\Vector{p}}$ are combined to produce the final pattern simulation. Molecular orientation that gives rise to anisotropy in the real-space structure of $\Vector{p}$ will produce correspondingly anisotropic patterns in the reciprocal space structure of the elements of $\Tilde{\Vector{p}}$, as illustrated for the disk morphology in \figref{fig: schematic}. There is typically a significant difference between the intensity of the polarized and depolarized scattering components such that the polarized scattering contributes most strongly to the sum, but the depolarized components remain essential for accurate simulation. This sum is not shown in \figref{fig: schematic} because a final step is required, the Ewalds projection.
 
 %can be written by taking $\Vector{r} \rightarrow \Vector{k}^\text{\textbf{out}}$ as
 
%  By making use of the first order Born approximation, the electric field $\Vector{e_s}$ resulting from X-ray interaction at location $\Vector{r}$, in the far-field limit is:
%  \begin{equation}
%       \lim_{r\to\infty} \Vector{e_s}(\Vector{r}) =\frac{e^{i\Vector{k}\cdot\Vector{r}}}{|r|} |r|^2\big(\Tensor{I} - \hat{r}\hat{r}\big)\cdot\Tilde{\Vector{p}}
%  \end{equation}
%  where: 
%  \begin{equation*}
%  \begin{split}
%      \Vector{r} \approx & ~ \Vector{k}^\text{\textbf{out}} = \Vector{k}^{\text{\textbf{in}}} + {\Vector{q}}
%     %  \Tensor{A} = & ~ |\Vector{r}|^2 (\Tensor{I} - \Vector{r}\Vector{r})
%     %  \textcolor{red}{\text{Compute}} \quad & X(\Vector{q}) = |\Tensor{A}\cdot\Vector{p}(\Vector{q})|^2
%  \end{split}{}
%  \end{equation*}{}

\item \textit{\textbf{Ewalds projection:}} The final step consists of projecting the differential scattering cross-section onto the Ewalds sphere to mimic the detector. For this step we will separately consider the elements of $\Vector{q}$ as $q_x$, $q_y$, and $q_z$. For each location on the detector given by $(q_x, q_y)$, we compute $q_z$ by evaluating
\begin{equation}
  \quad q_z  = -k_z^{\text{in}} + \sqrt{|\Vector{k}^{\text{\textbf{out}}}|^2 - (k_x^{\text{in}} + q_x)^2 - (k_y^{\text{in}} + q_y)^2}
    % |\Vector{k}^{\text{\textbf{out}}}| &  = |\Vector{k}^{\text{\textbf{in}}}| \\
    % \Vector{k}^{\text{\textbf{out}}} &  = \Vector{k}^{\text{\textbf{in}}}  + \Vector{q} 
\label{eq: Ewald}
\end{equation}{}
For real values of $q_z$, the detector image is given by interpolating $X(\Vector{q})$. Interpolation is needed because $q_z$ may not be an integer. We perform linear interpolation using the nearest integer neighbors. \figref{fig: schematic} shows the final scattering simulation after Ewalds projection of the $\Tilde{\Vector{p}}$ components also depicted.
\end{enumerate}

\section{Algorithm}
\label{sec: Algorithm}
The two criteria considered during algorithm design for \prsoxs{} simulation are the memory limitation on the GPU side and the communication time from central processing unit (CPU) to GPU. GPU architecture advancements have produced constant memory growth, but GPU memory remains much lower than its CPU counterpart. Additionally, data communication from CPU to GPU or vice versa remains a bottleneck. In this section, we describe the memory layout for the morphology and describe the two algorithms supported by our framework: a) \Algref{alg: prsoxs_comm_free} which minimizes the data movement from CPU to GPU but is memory intensive, especially for the larger number of material components; and b) \Algref{alg: prsoxs} which minimizes the memory footprint at the cost of communication between CPU and GPU.

\subsection{Memory layout for morphology} \label{sec:Morphology}
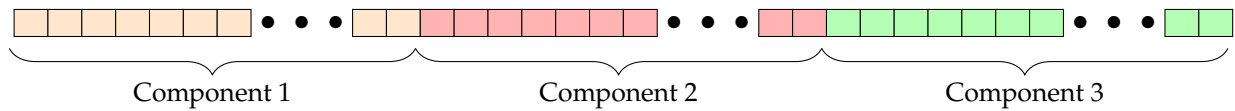
\begin{figure*}[h!]
\centering
    \begin{tikzpicture}[scale = 0.45]
    \foreach \x in {0,1,...,6}
        \draw[fill=orange!20] (\x,0) rectangle (\x+1,0.8);
    \foreach \x in {7,8,9}        
        \draw[draw = none,fill=none] (\x,0) rectangle (\x+1,0.8) node[pos=.5] {$\bullet$};
    \foreach \x in {10,11}                
        \draw[fill=orange!20] (\x,0) rectangle (\x+1,0.8); 
    \foreach \x in {12,13,...,18}        
        \draw[fill=red!30] (\x,0) rectangle (\x+1,0.8); 
    \foreach \x in {19,20,21}        
        \draw[draw = none,fill=none] (\x,0) rectangle (\x+1,0.8) node[pos=.5] {$\bullet$};  
    \foreach \x in {22,23}        
        \draw[fill=red!30] (\x,0) rectangle (\x+1,0.8);
    \foreach \x in {24,25,...,30}        
        \draw[fill=green!30] (\x,0) rectangle (\x+1,0.8);        
    \foreach \x in {31,32,33}
        \draw[draw = none,fill=none] (\x,0) rectangle (\x+1,0.8) node[pos=.5] {$\bullet$};
    \foreach \x in {34,35}
        \draw[fill=green!30] (\x,0) rectangle (\x+1,0.8);  
    \draw [decorate,decoration={brace,amplitude=10pt,mirror},xshift=-4pt,yshift=-10pt]
   (0,0) -- (12.0,0.0) node [black,below,midway,yshift=-0.3cm] {\footnotesize Component 1};
   \draw [decorate,decoration={brace,amplitude=10pt,mirror},xshift=-4pt,yshift=-10pt]
   (12,0) -- (24.0,0.0) node [black,below,midway,yshift=-0.3cm] {\footnotesize Component 2};
   \draw [decorate,decoration={brace,amplitude=10pt,mirror},xshift=-4pt,yshift=-10pt]
   (24,0) -- (36.0,0.0) node [black,below,midway,yshift=-0.3cm] {\footnotesize Component 3};
    \end{tikzpicture}
    \caption{Illustration of memory layout of morphology for a $c = 3$ component system, color coded for each component. The complete morphology is a 1D array of size $\text{N}_\text{x} \times \text{N}_\text{y} \times \text{N}_\text{z} \times \textsc{c}$. Each entry of morphology consists of a \texttt{Real4} entry.}
    \label{fig: memoryLayout}
\end{figure*}

The overall morphology is represented in memory as a 1D array of size $\mathrm{N_x} \times \mathrm{N_y} \times \mathrm{N_z} \times c$. Each entry of this 1D array consists of a \texttt{Real4} \footnote{\texttt{Real4} represents \texttt{float4}, a single-precision floating-point number, or \texttt{double4}, a double-precision floating-point number, depending upon type of compilation} data type representing the 4 components ($v_{frac},s,\phi,\theta$). \figref{fig: memoryLayout} shows the memory layout of morphology for \prsoxs{} simulation. Using a 1D array ensures that only a single \texttt{cudaMemcpy} instruction is needed to load from CPU to GPU memory. The use of \texttt{Real4} datatype ensures vectorized load from global memory of GPU to local memory. Additionally, this memory layout -- of striding through voxels first before striding through components -- ensures the best utilization of the load bandwidth from global memory to local memory. %This is mainly because the neighboring threads in a block load the consecutive entries from the morphology at a particular instruction.

The memory layout allows for additional computational gains during the averaging process. An earlier algorithm by \citet{gann2016origins} relied on rotating the material, keeping $\Vector{e}$ fixed, in order to compute the average intensity on the detector. This step is computationally expensive, especially for 3D morphologies, where we would need to rotate $\mathrm{N_z}$ channels of $\mathrm{N_x} \times \mathrm{N_y}$ voxelated morphologies. In this work, we reformulated the algorithm to rotate $\Vector{e}$ while keeping the material fixed. This requires all the vectors $\Vector{p}$ to be evaluated in the rotated $\Vector{e}$ reference frame rather than the fixed material reference frame. Additionally, we rotate the detector coordinates at the last step to average the resulting intensity. The transfer of computation from the material to $\Vector{e}$ reference frame makes the algorithm computationally efficient and  GPU friendly.

\subsection{Communication Minimization (\texorpdfstring{\Algref{alg: prsoxs_comm_free}}{})} 

\begin{algorithm}[b!]
  \caption{\prsoxs~ simulation : Communication minimization}
  \label{alg: prsoxs_comm_free}
  \footnotesize
  \begin{algorithmic}[1]
    \Require $\Vector{M}$ morphology information; \textbf{E}: Energy List ; $\mathrm{\textbf{E}}_{\mathrm{\textbf{Angle}}}$[start,increment,End]: rotation angle for  $\Vector{e}$;  $\Tensor{N}^j$: Refractive Index for each material at all energy
    \Ensure 2D RSoXS pattern
     \State $\Vector{M}^\mathrm{GPU}$ $\leftarrow$ $\Vector{M}^\mathrm{CPU}$ \Comment{Copy from CPU to GPU} 
     \For{i $\leftarrow 1 \cdots \mathrm{\textbf{len}}(E)$}
     \State $\mathrm{Scatter}_{\mathrm{Avg}}[i] \leftarrow 0$
     \For{$e_{Angle} \in \mathrm{\textbf{E}}_{\mathrm{\textbf{Angle}}}$}
     \State Compute $\Vector{e_{rot}}$ \Comment{Rotate electric field using rotation matrix $\Tensor{U}$}
     \For{each voxel}
     \State Compute $\Tensor{N_{eff}}^j$ \Comment{\eqnref{eq: ncorr}}
    \State Compute $\Tensor{N_{rot}}^j$ \Comment{\eqnref{eq: nrot}}
     \State Compute $\Vector{p}$
      \Comment{\eqnref{eq: polarization}}
     \EndFor
     \State $\Vector{p} \leftarrow \Tensor{U}^T\cdot\Vector{p}$
     \State $\Tilde{\Vector{p}} \leftarrow$  FFT($\Vector{p}$))
     \Comment{in-place FFT transform}
     \For{each pixel $(q_x,q_y)$ in 2D}
     \State Compute $q_z$ for projection \Comment{\eqnref{eq: Ewald}}
     \State Ewald[$q_x,q_y$] $\leftarrow$  $\Vector{X}(q_z)$ \Comment{\eqnref{eq: 3Dscatter}}
     \EndFor
     \EndFor
     \State $\mathrm{Ewald}_{\mathrm{Avg}}[i] \leftarrow \mathrm{Ewald}_{\mathrm{Avg}}[i]$/numRotation \Comment{Average the contribution over all the $\Vector{e}$ rotations}
     \EndFor
     \State \Return $\mathrm{Ewald}_{\mathrm{Avg}}$
  \end{algorithmic}
\end{algorithm}
This algorithm relies on copying all the morphology information once from CPU to GPU at the start of the computation, which is then utilized for all the subsequent computations. Once this copy is performed, no communication is needed for the next computation steps. We perform the polarization computation $\Vector{p}$, given by \eqnref{eq: polarization}. As discussed in the previous section, the memory layout for the vector morphology allows us to achieve maximum bandwidth mainly because all subsequent threads within the block try to load the nearby memory. Additionally, packing the data as \texttt{Real4} allows us to perform vectorized load from global memory to local thread memory. To efficiently utilize the available resources, we use streams to compute the FFT of the polarization vector. In particular, we use 3 streams, one for each of $p_x,p_y$ and $p_z$. We then compute the $q_z$ position for a given value of $(q_x, q_y)$ 2D pixel, given by \eqnref{eq: Ewald}. We note that we only compute $X(q_z)$ for the pixels participating in 3D Ewald's projection. This helps to eliminate the memory requirement to store a 3D vector for $X(\Vector{q})$. Finally, the averaged result (averaged across a range of rotation angles of $\Vector{e}$) is transferred from GPU to CPU. \Tabref{tab:memoryAlg1} shows the memory requirement for \prsoxs{} simulation. One potential drawback of this approach is the overall memory requirement. We can see that overall memory requirement grows linearly with the number of materials. Memory requirements are dependent on the resolution and number of materials per model, and can range from less than 1~GB to approaching or exceeding the $\approx$~48~GB memory limit of current-generation CUDA GPUs.

\begin{table*}[b!]
    \footnotesize
    \centering
    \setlength{\tabcolsep}{12pt} % Default value: 6pt
    \caption{Memory requirement for various steps during \prsoxs{} computation for ~\Algref{alg: prsoxs_comm_free}}
    \label{tab:memoryAlg1}
    \begin{tabular}{ccccc}
    \hline
    Algorithm & Variable & Data Type & Size  & Total Size  \\ \hline
    \multirow{9}{*}{\begin{tabular}[c]{@{}c@{}}\prsoxs{}\\ (Algorithm~\ref{alg: prsoxs_comm_free}) \end{tabular}} 
    & {$\Vector{M}$} & {Real} & $4c$$ (n_x n_y n_z)$  & $4c$$ (n_x n_y n_z)$ \\
    & ${p_x}$ & Complex & $(n_x n_y n_z)$ &  $2(n_x n_y n_z)$ \\
    & ${p_y}$ & Complex & $(n_x n_y n_z)$ &  $2(n_x n_y n_z)$ \\
    & ${p_z}$ & Complex & $(n_x n_y n_z)$ &  $2(n_x n_y n_z)$ \\
    & Ewald & Real & $(n_x n_y)$ &   $(n_x n_y)$ \\
    & Ewald$_{\mathrm{Avg}}$ & Real & $(n_x n_y)$ &   $(n_x n_y)$ \\ \cline{2-5}
    & {Total} &  &   & $(4c+6)(n_x n_y n_z) + 2(n_x n_y)$ \\  \hline
    \end{tabular}
\end{table*}
\subsection{Memory Minimization (\texorpdfstring{\Algref{alg: prsoxs}}{})}
\begin{algorithm}[t!]
  \caption{\prsoxs~ simulation : Memory minimization}
  \label{alg: prsoxs}
  \footnotesize
  \begin{algorithmic}[1]
    \Require $\Vector{M}$ morphology information; \textbf{E}: Energy List ; $\mathrm{\textbf{E}}_{\mathrm{\textbf{Angle}}}$[start,increment,end]: rotation angle for  $\Vector{e}$;  $\Tensor{N}^j$: Refractive Index for each material at all energy
    \Ensure 2D RSoXS pattern
     
     \For{i $\leftarrow 1 \cdots \mathrm{\textbf{len}}(E)$}
     \State $\Tensor{N_{t}}^\mathrm{GPU} \leftarrow \textsc{NrComputation} (\Vector{M},E,\Tensor{N}^j)$ \Comment{Algorithm: \ref{alg: localNr}}
     \State $\mathrm{Scatter}_{\mathrm{Avg}}[i] \leftarrow 0$
     \For{$e_{Angle} \in \mathrm{\textbf{E}}_{\mathrm{\textbf{Angle}}}$}
     \State Compute $\Vector{e_{rot}}$ \Comment{Rotate electric field using rotation matrix $\Tensor{U}$}
     
     \For{each voxel}
     \State Compute $\Vector{p}(\Vector{x}) \leftarrow \Tensor{N_t}\cdot \Vector{e_{rot}}$ \Comment{\eqnref{eq: polarization}}
     \EndFor
     \State $\Vector{p} \leftarrow \Tensor{U}^T\cdot\Vector{p}$
     \State $\Tilde{\Vector{p}} \leftarrow$  FFT($\Vector{p}$))
     \Comment{in-place FFT transform}
     \For{each pixel $(q_x,q_y)$ in 2D}
     \State Compute $q_z$ for projection \Comment{\eqnref{eq: Ewald}}
     \State Ewald[$q_x,q_y$] $\leftarrow$  $\Vector{X}(q_z$) \Comment{\eqnref{eq: 3Dscatter}}
     \EndFor
     \EndFor
     \State $\mathrm{Ewald}_{\mathrm{Avg}}[i] \leftarrow \mathrm{Ewald}_{\mathrm{Avg}}[i]$/numRotation \Comment{Average the contribution over all the $\Vector{e}$ rotations}
     \EndFor
     \State \Return $\mathrm{Ewald}_{\mathrm{Avg}}$
  \end{algorithmic}
\end{algorithm}
\begin{algorithm}[ht!]
  \caption{\textsc{NrComputation:} Local Refractive Index computation}
  \label{alg: localNr}
  \footnotesize
  \begin{algorithmic}[1]
    \Require $\Vector{M}$: morphology information; E: Energy ; $\Tensor{N}^j$: Refractive Index for each material for energy E
    \Ensure $\Tensor{N_\mathrm{t}}^\mathrm{CPU} = (\Tensor{N_r}^j:\Tensor{N_r}^j - \Tensor{I})/(4\pi)$ for energy E
    \State \textsc{cudaMalloc} $\Vector{M}$ ,$\Tensor{N_{i}} $ \Comment{Allocate GPU memory}
    \For {each voxel}
    \State $\Tensor{N_{t}} \leftarrow 0$ \Comment{Initialize}
    \For{j $\leftarrow 1 \cdots c$} \Comment{Loop over components}
    \State \textsc{memCopy} $\Vector{M}^j$ : CPU $\rightarrow$ GPU \Comment{Copy from CPU to GPU component by component}
    \State Compute $\Tensor{N_{eff}}^j$ \Comment{\eqnref{eq: ncorr}}
    \State Compute $\Tensor{N_{rot}}^j$ \Comment{\eqnref{eq: nrot}}
    \State $\Tensor{N_t} \leftarrow \Tensor{N_t} + (\Tensor{N_r}^j:\Tensor{N_r}^j - \Tensor{I})/(4\pi)$  
    \EndFor
    \EndFor
    \State \textsc{cudaFree $\Vector{X}$} \Comment{Free CUDA memory for morphology}
    \State \Return $\Tensor{N_{t}}^\mathrm{CPU}$
  \end{algorithmic}
\end{algorithm}
\begin{table*}[t!]
    \footnotesize
    \centering
    \setlength\extrarowheight{3pt}
    \newcommand{\tabincell}[2]{\begin{tabular}{@{}#1@{}}#2\end{tabular}}
    \setlength{\tabcolsep}{12pt} % Default value: 6pt
    \caption{Memory requirement for various phases during \prsoxs{} computation for ~\Algref{alg: prsoxs}}
    \label{tab:memory}
    \begin{tabular}{ccccc}
    \hline
    Algorithm & Variable & Data Type & Size  & Total Size  \\ \hline
    \multirow{9}{*}{\begin{tabular}[c]{@{}c@{}}~\prsoxs{}\\ (Algorithm~\ref{alg: prsoxs}) \end{tabular}} & $\Tensor{N_t}$ & Complex & $6(n_x n_y n_z)$  & $12(n_x n_y n_z)$ \\
     & ${p_x}$ & Complex & $(n_x n_y n_z)$ &  $2(n_x n_y n_z)$ \\
     & ${p_y}$ & Complex & $(n_x n_y n_z)$ &  $2(n_x n_y n_z)$ \\
     & ${p_z}$ & Complex & $(n_x n_y n_z)$ &  $2(n_x n_y n_z)$ \\
     & Ewald & Real & $(n_x n_y)$ &   $(n_x n_y)$ \\ 
     & Ewald$_{\mathrm{Avg}}$ & Real & $(n_x n_y)$ &   $(n_x n_y)$ \\ \cline{2-5}
    & {Total} &  &   & $18(n_x n_y n_z) + 2(n_x n_y)$ \\  \hline
    %%%%%%%%%%%%%%%%%%%%%%%%%%%%%%%%%%%%%%%%%%%%%%%%%%%%%%%%%%%%%%%%%%%%%%%
    \multirow{3}{*}{\begin{tabular}[c]{@{}c@{}}\textsc{NrComputation}\\ (Algorithm~\ref{alg: localNr}) \end{tabular}} & {$\Vector{M}$} & {Real} & $4$c$ (n_x n_y n_z)$  & $4c$$ (n_x n_y n_z)$ \\
    &{$\Tensor{N_t}$} & {Complex} & $6(n_x n_y n_z)$  & $12(n_x n_y n_z)$ \\
    %  &  &  & $6(n_x n_y n_z)$ & biaxial & $12(n_x n_y n_z)$ \\
     \cline{2-5}
    & \tabincell{c}{Total\\(non - stream)} &  & & $(4c + 12)(n_x n_y n_z)$ \\ 
    & \tabincell{c}{Total\\(stream)} &  & & $(4 + 12)(n_x n_y n_z)$ \\ 
    \hline
    \end{tabular}
\end{table*}

 Analysis of the steps detailed in \secref{sec: MathSteps} indicates that morphology inputs are only required during the computation of polarization $\Vector{p}$, \eqnref{eq: polarization}. The main idea of this algorithm is to precompute a precursor of $\Vector{p}$ for a given energy and use it for all subsequent computation (across multiple rotations of $\Vector{e}$). The pre-computation stage is shown by \Algref{alg: localNr}, which computes an intermediate tensor $\Tensor{N_t}$ (which is defined as $\sum_{j=1}^c (\Tensor{N_r}^j:\Tensor{N_r}^j - \Tensor{I})/(4\pi)$, see \eqnref{eq: polarization}). The computation in this step is embarrassingly parallel and can be computed per voxel independently. Therefore, even if the complete memory required does not fit on the GPU, we can asynchronously stream the required data to and from CPU to GPU. In particular, we stream the data per material from CPU to GPU. So, the memory requirement during this stage drops from $(4c + 12)(n_xn_yn_z)$ to $16(n_xn_yn_z)$. The streaming helps to overlap computation with communication and hides the latency.

Once $\Tensor{N_t}$'s are computed, these values are subsequently used for the  \prsoxs{} simulation in a similar way as in \Algref{alg: prsoxs_comm_free}. \Tabref{tab:memory} shows the memory requirement for the different steps. The memory requirement for the main stage is independent of the number of materials and requires less memory compared to \Algref{alg: prsoxs_comm_free} for $c\geq 3$. This is an important consideration, especially when we consider multi-component chemical systems. Finally, we exploit the symmetric structure of $\Tensor{N_t}$ to further minimize the number of computations required. While $\Tensor{N_t}$ contains 9 entries ($3 \times 3$ matrix), only 6 of these entries are unique. 

\begin{remark}
We note that further optimization is possible in terms of memory requirement. Theoretically only $6(n_x n_y n_z) + 2 (n_x n_y)$ (3 $\Vector{p}$ vectors and 2 vectors for Ewald and $\mathrm{Ewald}_\mathrm{Avg}$) units of memory is required for \prsoxs{} computation. All the other information can be communicated from CPU to GPU in a streamed fashion. But achieving this theoretical bound would imply a lot of communication overhead with $\Vector{M}$ or $\Tensor{N_t}$ being communicated from CPU to GPU for each rotation of $\Vector{e}$ field. For most of our use cases, we find that the memory available on current GPUs, like the \textsc{Nvidia Voltas V100}, is sufficient for carrying out the computation using \Algref{alg: prsoxs_comm_free} with \Algref{alg: prsoxs} needed in some extreme cases.
\end{remark}

\section{Results: Validation Cases}
\label{sec:results}
We comprehensively verify and validate CyRSoXS by comparing against an array of benchmarks. This includes three test cases with analytical scattering expressions, and one validation case consisting of comparisons against results within an earlier framework~\citep{gann2016origins}.

\subsection{Form factor scattering test}
\par A simple validation case is for form factor scattering, in which scattering results purely from the shape of a particle. We specifically test the form factor scattering of a sphere. We consider two cases: a) 2D projection of a sphere, and b) 3D sphere, and we compare the results of CyRSoXS to analytical expression results. The analytical expression for form factor scattering of a sphere is given by:
\begin{equation}
    I(q)={\frac {\rm {scale}}{V}}\left[{\frac {3V(\Delta \rho )(\sin(qr)-qr\cos(qr))}{(qr)^{3}}}\right]^{2}
    \label{eq:FoamFactor}
\end{equation}
where, scale is the intensity scaling, $r$ is the sphere radius (in \si{\angstrom}), $\Delta \rho$ is the scattering contrast (in \si{\angstrom}$^{-2}$).

\subsubsection{2D projection of a sphere}
\begin{figure}[t!]
    \centering
    \includegraphics[width=0.5\linewidth]{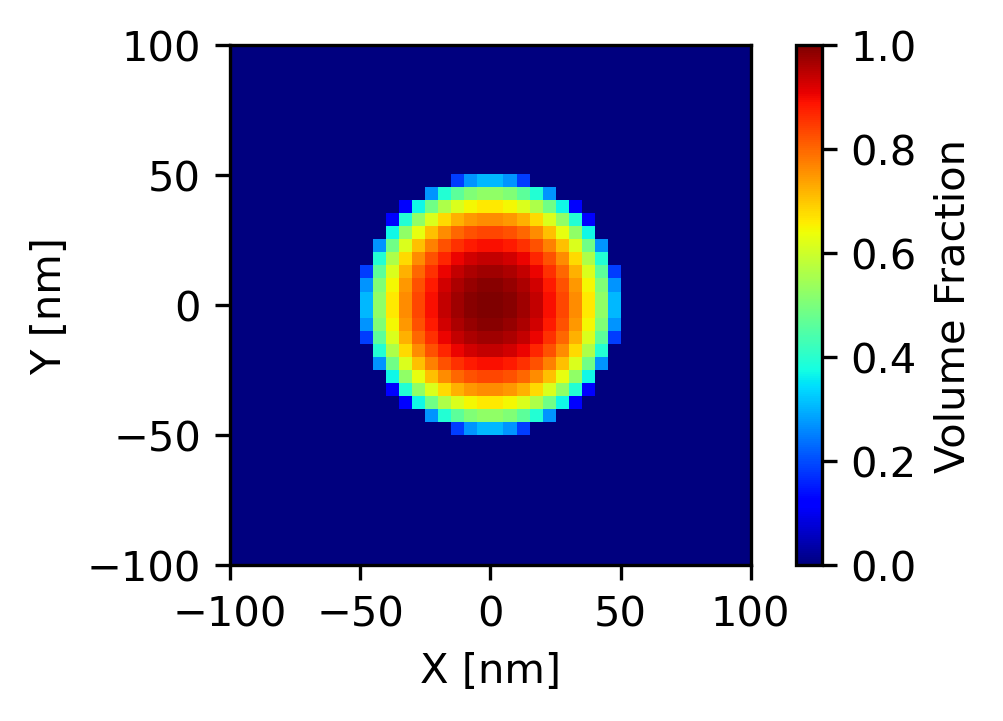}
    \caption{A zoomed-in view of the 2D projected sphere}
    \label{fig:2DCircleSetup}
\end{figure}

As a first test case, we consider a 2D projection of a sphere of radius 50 nm placed in the center of the domain. For this test, the sphere is composed of amorphous polyethylene in a surrounding medium of vacuum.~\figref{fig:2DCircleSetup} illustrates the zoomed view of domain setup for the test case. The whole domain is discretized using $2048 \times 2048 \times 1$ voxels, with each voxel representing a $5 \times 5 \times 5$ nm$^3$ physical volume. \figref{fig:2DCircleSetup} shows a zoomed view near the center of the circle. The scattering profile for this morphology was simulated from (270 to 310) eV, using tabulated optical constants of polyethylene~\citep{eliottGithub} for the projected sphere and vacuum outside. 

\begin{figure*}[t!]
    \centering
    \includegraphics[width=0.99\linewidth]{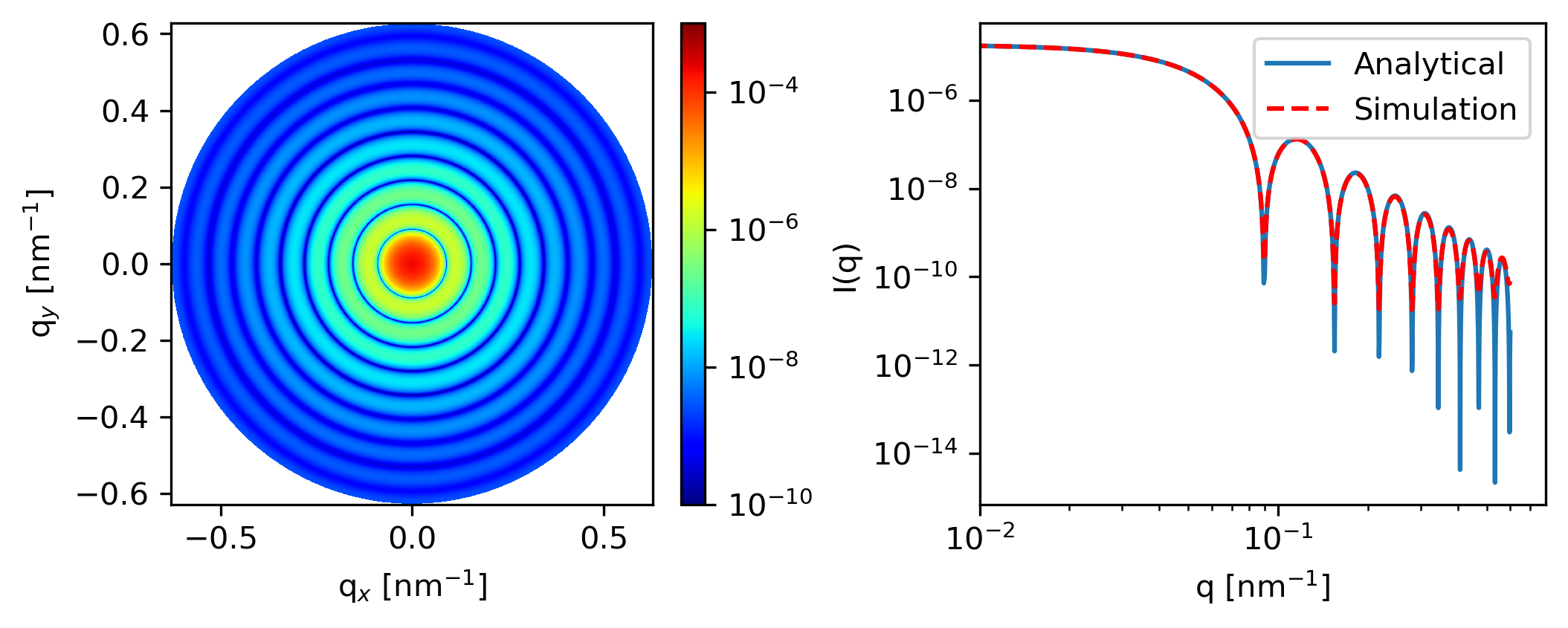}
    \caption{Results of the projected sphere case and comparison with analytical solution.}
    \label{fig:2DCircleResults}
\end{figure*}

\begin{figure}[t!]
    \centering
    \includegraphics[width=0.5\linewidth]{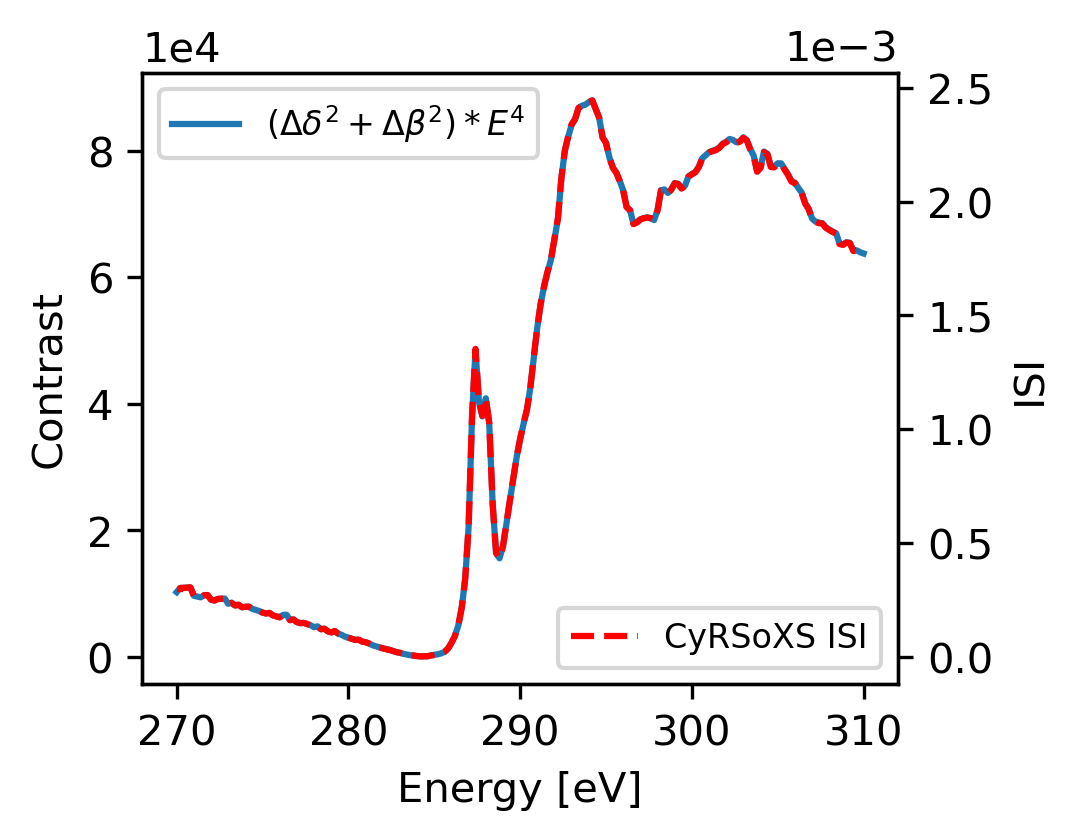}
    \caption{The simulated ISI alongside the theoretical energy dependence.}
    \label{fig:2DEDependence}
\end{figure}

\figref{fig:2DCircleResults} shows the result of the 2D projected sphere validation case at 285 eV. Linecuts of the analytical and simulated data are plotted in Figure 5b and show excellent agreement. To validate the energy dependence of the \prsoxs{} simulation, we calculate the integrated scattering intensity (ISI) across the range of simulated energy values. \figref{fig:2DEDependence} plots the simulated ISI alongside the theoretical energy dependence given by the analytical expressions provided in Tatchev\citep{tatchev2010multiphase}, computed for this specific dielectric function by us:
\begin{equation}
    (\Delta \delta^2 + \Delta \beta^2)*E^4
\end{equation}
While on different absolute scales, the theoretical and simulated energy dependence show commensurate relative scaling indicating we are capturing the correct physics in our scattering model.

\subsubsection{3D sphere test}

\begin{figure*}[t!]
    \centering
    \begin{subfigure}{0.48\linewidth}
        \includegraphics[trim = 0 0 300 0, clip, height=2.5in]{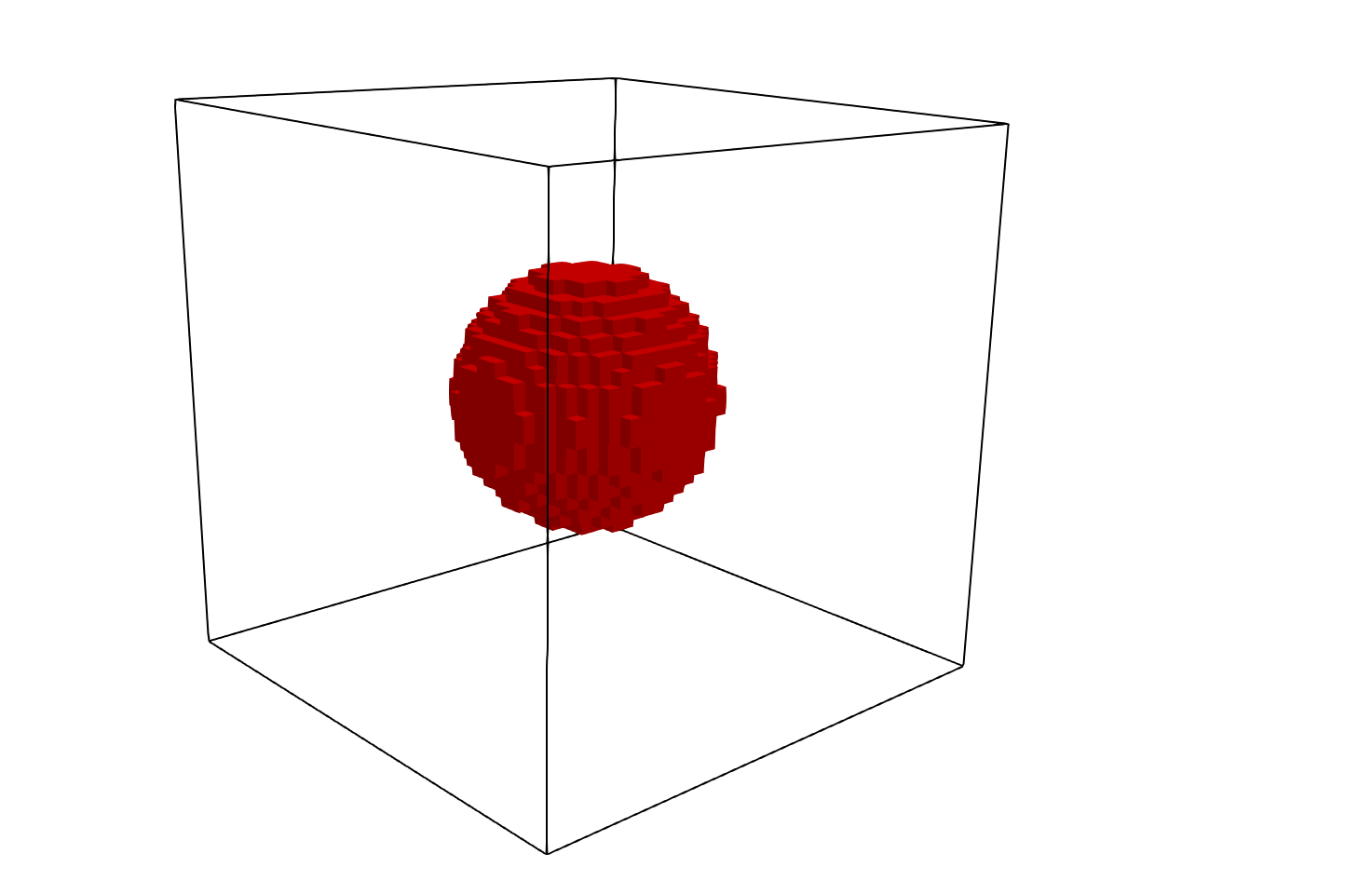}
        \caption{3D sphere}
        \label{fig:3DSphere}
    \end{subfigure}
    \hfill
    \begin{subfigure}{0.48\linewidth}
        \includegraphics[trim=100 0 150 0,clip,height=2.5in]{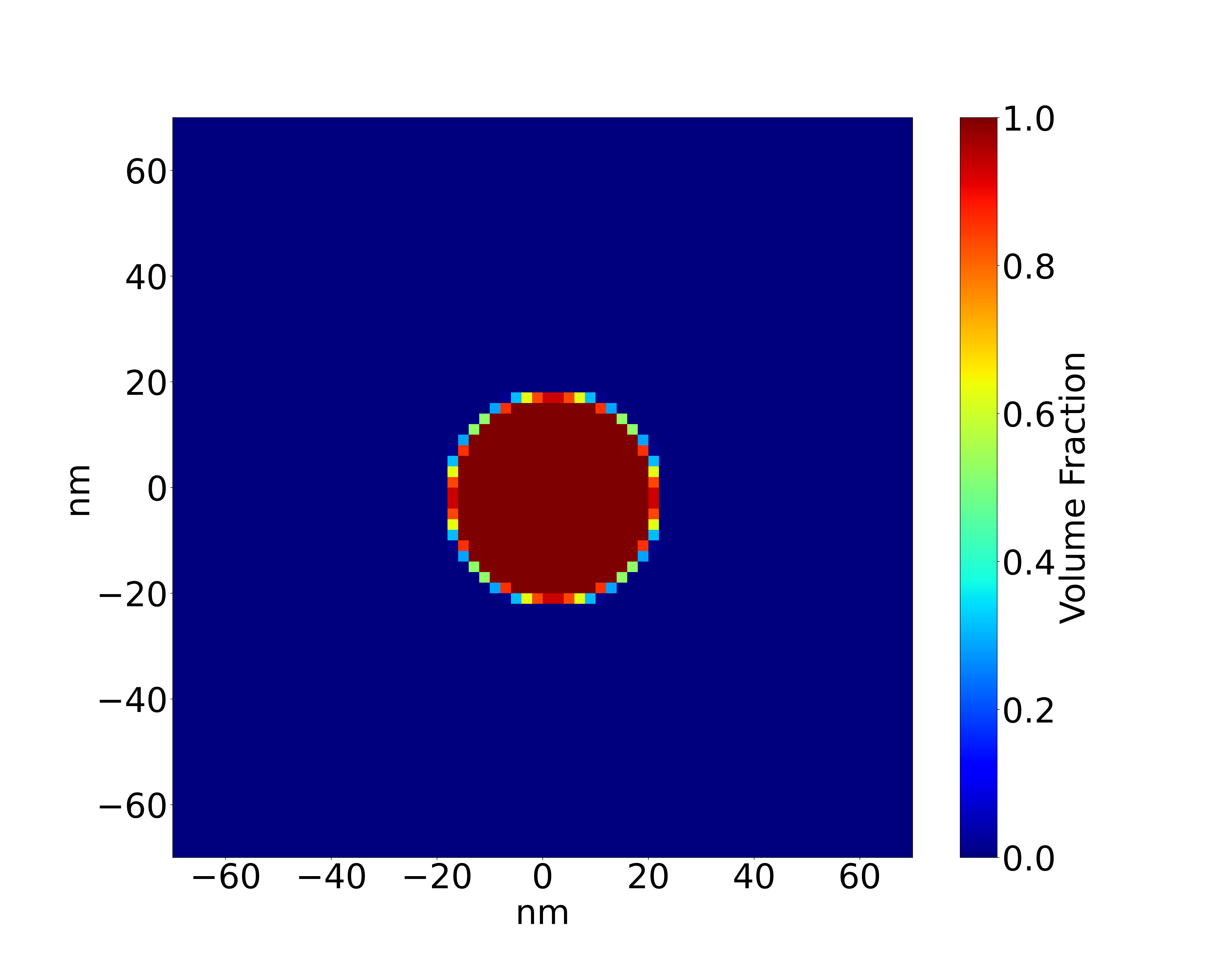}
        \caption{2D mid plane slice}
        \label{fig:3DSphereSlice}
    \end{subfigure}
    \caption{Domain for the sphere validation test (not to scale) and a 2D slice along the mid plane. }
    \label{fig:SphereValidationCase}
\end{figure*}

\figref{fig:SphereValidationCase} shows the 3D sphere test domain along with a 2D slice of the sphere mid-plane. The morphology consists of $128 \times 2048 \times 2048$ voxels, where each voxel is $5 \times 5 \times 5$ nm$^3$. A 3D sphere of radius 50 nm is placed at the center. The simulation was carried out at 285 eV, using tabulated polyethylene optical constants for the sphere and vacuum for the surrounding matrix.

\begin{figure*}[t!]
    \centering
    \includegraphics[width=0.99\linewidth]{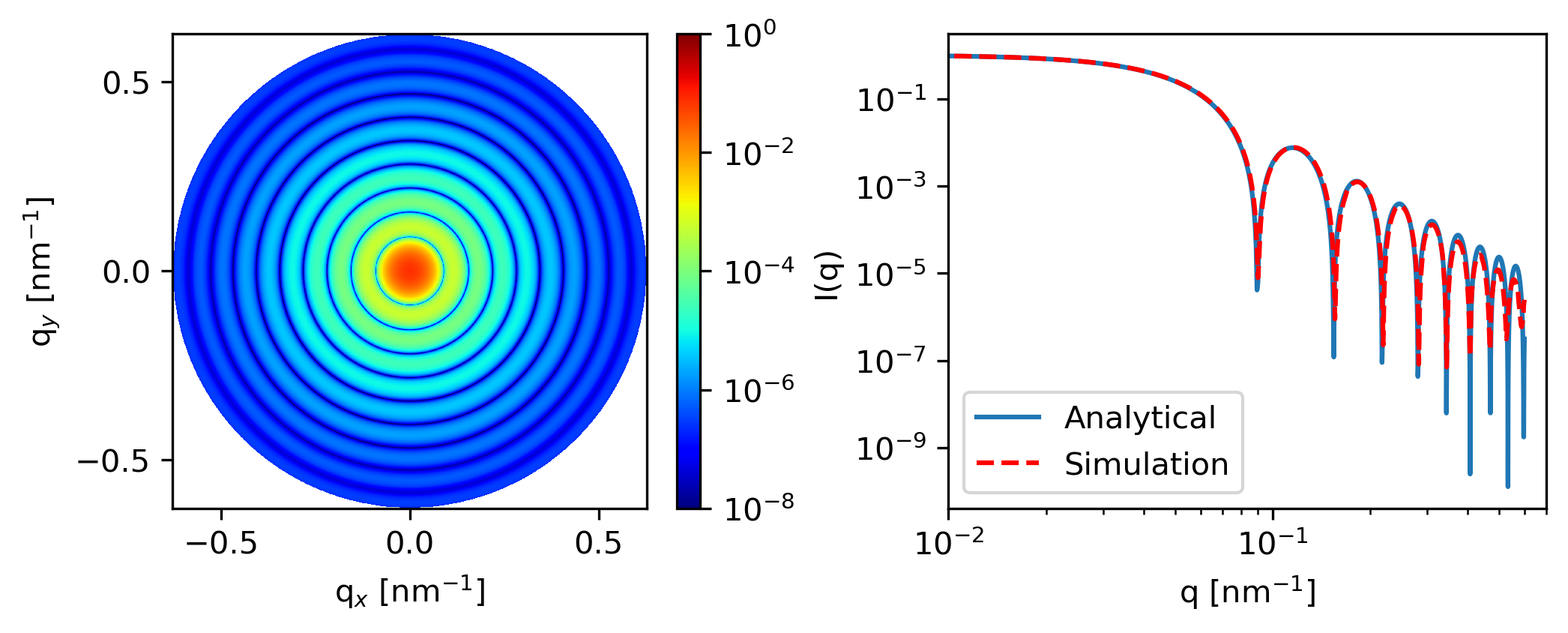}
    \caption{Results of the 3D sphere case and comparison with analytical solution.}
    \label{fig:3DSphereResults}
\end{figure*}

\figref{fig:3DSphereResults}a shows the 2D scattering pattern and \figref{fig:3DSphereResults}b compares the 1D analytical expression for a sphere with the azimuthally integrated data from \figref{fig:3DSphereResults}a. The analytical and simulation data were both normalized to 1 at q = 1e-2. We see an excellent comparison between the analytical and simulated results. The minor discrepancy between the simulation and analytical results at higher q values can be attributed to the finite discretization of the sphere and voxel size.

\subsection{Periodic structure test}
\par Extending beyond form factor scattering, many materials studied with X-ray scattering techniques exhibit periodic structures, which result in Bragg diffraction: constructive interference of the scattered X-rays produces sharp peaks at locations corresponding to the periodic spacing. Materials of this nature that have been studied with RSoXS include block copolymers~\citep{wang2011defining, virgili2007analysis} and patterned thin films~\citep{freychet2018using}. Voxelized representations will approximate the spacings and shapes of real morphologies. We perform two validation cases that reflect this periodic arrangement of structures: a) 2D hexagonal packed lattice; and b) grating test.
\subsubsection{Circle on hexagonal lattice}
 We first consider an arrangement of circular domains on a 2D hexagonal lattice. This morphology is representative of hexagonally-packed cylinders, a common block copolymer morphology. We consider the cylinders to be oriented parallel to the X-ray beam.
 
\begin{figure}[h!]
    \centering
    \includegraphics[trim = 0 100 0 180,clip,width=0.5\textwidth]{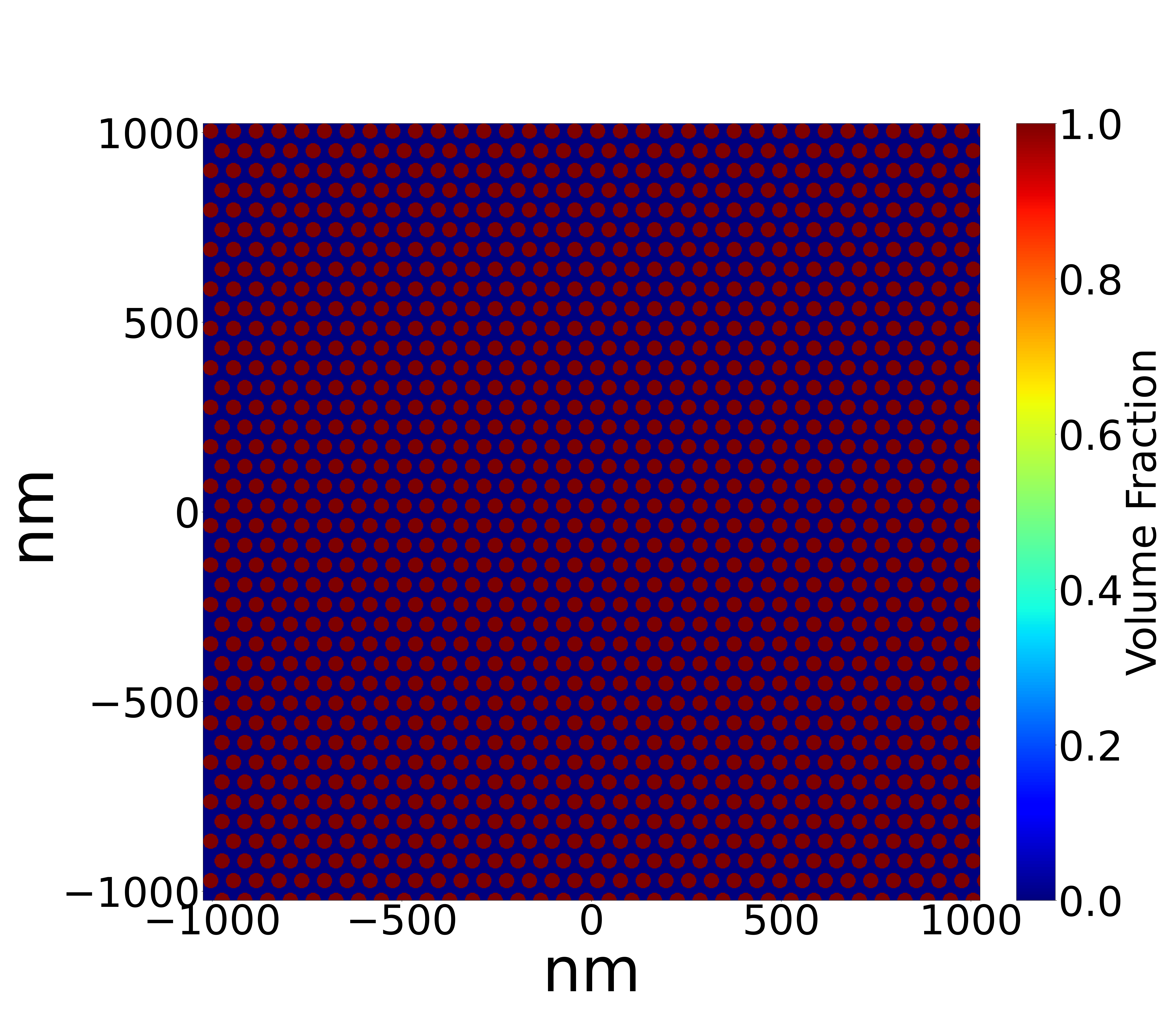}
    \caption{Volume fraction map of PEOlig for the hexagonal lattice. The dark blue region represents vacuum.}
    \label{fig:hexLatticeSetup}
\end{figure}

\begin{figure}[t!]
    \centering
    \includegraphics[width=0.99\linewidth]{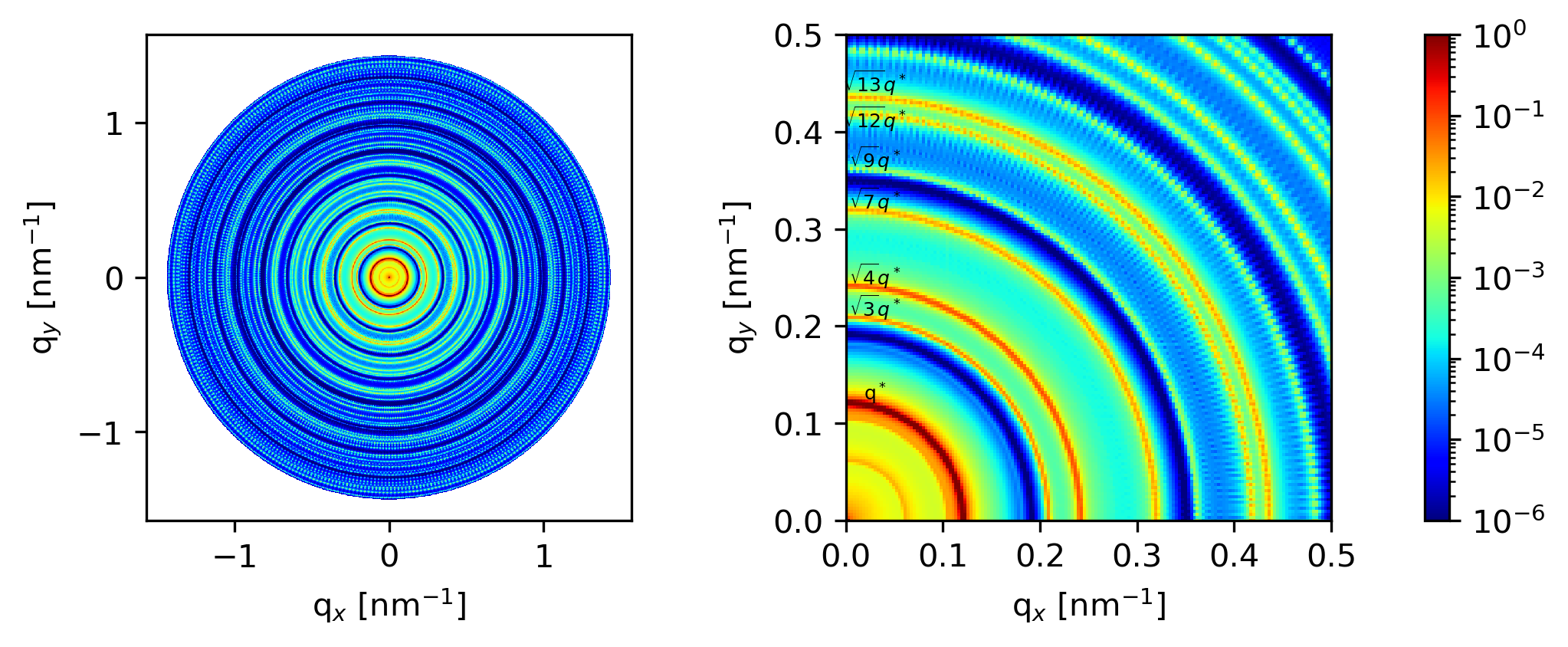}
    \caption{Hexagonal lattice validation case. Contours of I($\Vector{q}$) with the corresponding peak locations.}
    \label{fig:HexagonalLattice}
\end{figure}

\begin{figure*}[t!]
    \centering
    \includegraphics[width=0.6\textwidth]{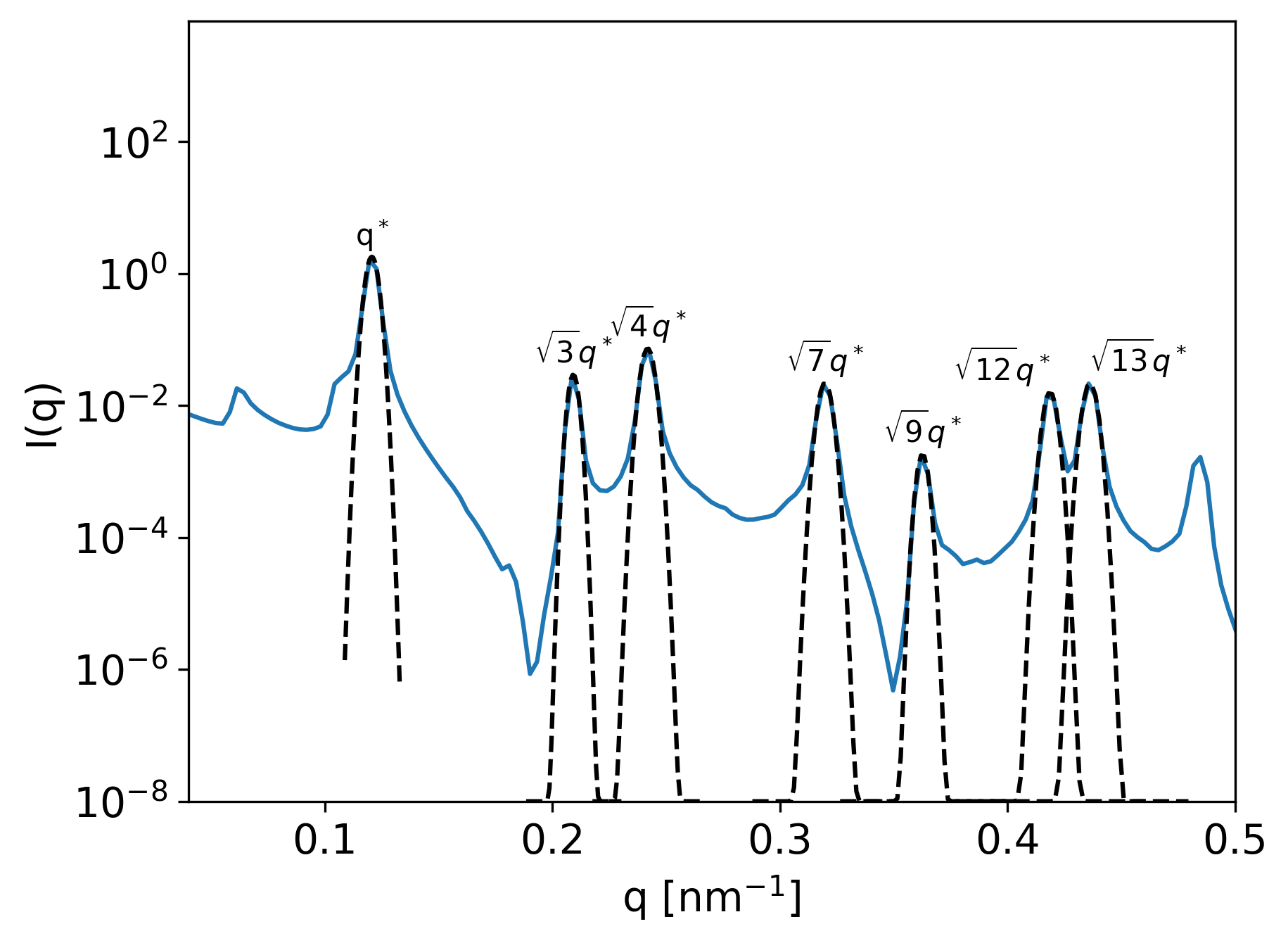}
    \caption{1D simulated diffraction pattern with the analytical peak locations marked.}
    \label{fig:HexagonalLineCuts}
\end{figure*}

\figref{fig:HexagonalLattice} shows the 2D scattering pattern output from Cy-RSoXS. Given the target lattice spacing of the input morphology, we observe Bragg peaks at the expected locations ($q^{*}, \sqrt{3}q^{*}, \sqrt{4}q^{*}, \sqrt{7}q^{*}, \sqrt{9}q^{*},$ and so on). \figref{fig:HexagonalLineCuts} shows the azimuthally integrated scattering intensity plotted versus q, with the first 7 Bragg peaks labeled. There is perfect agreement between the analytical and simulated peak locations. We do observe some non-peak background features with low intensity that originate from the finite size of the model and voxel-level discretization effects. Such artifacts can be further reduced by using larger and/or higher-resolution models, models that contain realistic structural defects, and models with periodic boundary conditions.

\subsubsection{Grating test}

\begin{figure}[b!]
    \centering
    \includegraphics[width=3.25in]{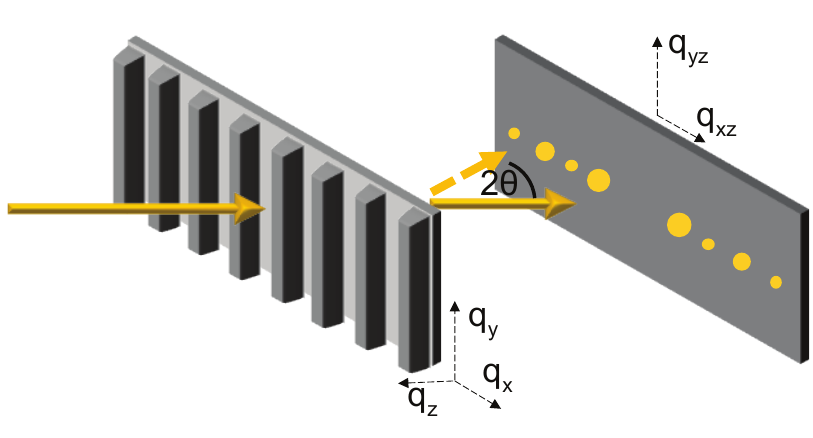}
    \caption{Setup of the line grating simulation}
    \label{fig:Grating-setup}
\end{figure}

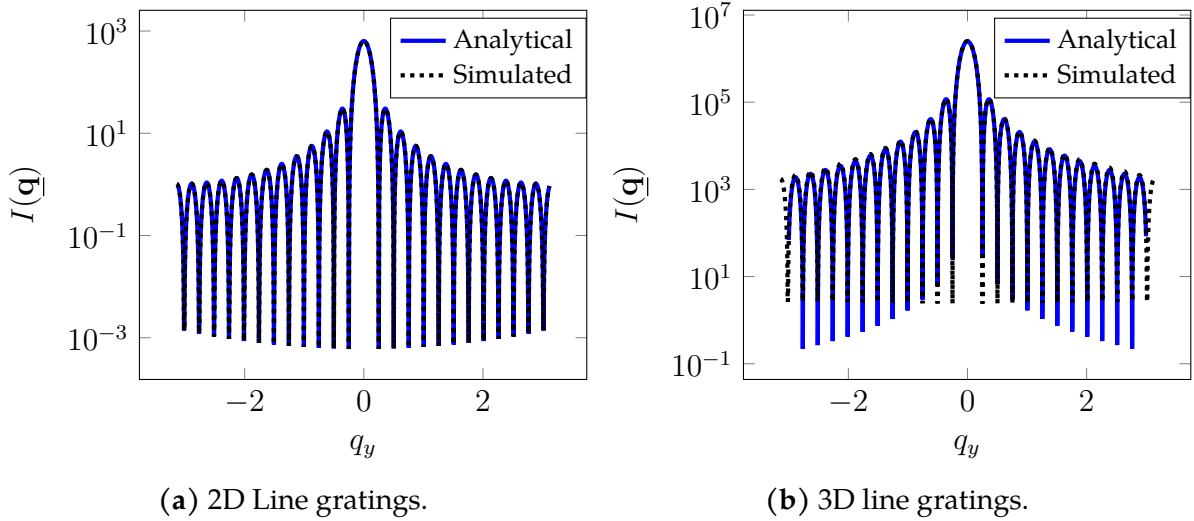
\begin{figure}[t!]
    \begin{subfigure}[b]{0.48\textwidth}
    \centering
    \begin{tikzpicture}
    \begin{semilogyaxis}[
      %width=0.6\linewidth, % Scale the plot to \linewidth
      width=0.95\textwidth,
      %height=0.2\textwidth,
      %grid=major, % Display a grid
      %grid style={line width=.1pt, draw=gray!50}, % Set the style
      xlabel=$q_y$, % Set the labels
      ylabel=$I(\Vector{q})$,
    %   xmin=0,
    %   xmax=3.5,
    %   ymax = 1.1,
    %   legend style={at={(1.2,0.7)},anchor=north west,legend columns=1}, % Put the legend below the plot
      x tick label style={rotate=0,anchor=north}, % Display labels sideways
      legend style={at={(0.78,0.98)},anchor= north,legend columns=1}, % Put the legend below the plot
    ]
           \addplot[color=blue,ultra thick,no marks]
       table [col sep=comma, x=Qy, y=Intensity]{Data/Gratings/2DAnalytical.txt};
%    \addplot [no marks] gnuplot [raw gnuplot, color=blue,ultra thick] {
%      set datafile separator ",";
%      plot "Data/Gratings/2DAnalytical.txt" using
%      1:2 with lines
%     };
 \addplot[color=black,ultra thick,dotted,no marks]
table [col sep=space, x=Qy, y=Intensity]{Data/Gratings/BraggData2D.txt};
%     \addplot [no marks] gnuplot [raw gnuplot, color=black,ultra thick, dotted] {
%      set datafile separator " ";
%      plot "Data/Gratings/BraggData2D.txt" using
%      1:2 with lines
%     };
     \legend{\footnotesize Analytical ,\footnotesize Simulated}
     \end{semilogyaxis}
     \end{tikzpicture}
     \caption{2D Line gratings.}
     \label{fig: 2DBraggs}
     \end{subfigure}
     \vspace{0.02\textwidth}
     \begin{subfigure}[b]{0.48\textwidth}
     \centering
     \begin{tikzpicture}
     \begin{semilogyaxis}[
          %width=0.6\linewidth, % Scale the plot to \linewidth
          width=0.95\textwidth,
          %height=0.2\textwidth,
          %grid=major, % Display a grid
          %grid style={line width=.1pt, draw=gray!50}, % Set the style
          xlabel=$q_y$, % Set the labels
          ylabel=$I(\Vector{q})$,
        %   xmin=0,
        %   xmax=3.5,
        %   ymax = 1.1,
        %   legend style={at={(1.2,0.7)},anchor=north west,legend columns=1}, % Put the legend below the plot
          x tick label style={rotate=0,anchor=north}, % Display labels sideways
          legend style={at={(0.78,0.98)},anchor= north,legend columns=1}, % Put the legend below the plot
     ]
       \addplot[color=blue,ultra thick,no marks]
     table [col sep=comma, x=Qy, y=Intensity]{Data/Gratings/analyticalBragg3D.txt};
%    \addplot [no marks] gnuplot [raw gnuplot, color=blue,ultra thick] {
%    set datafile separator ",";
%    plot "Data/Gratings/analyticalBragg3D.txt" using
%    1:2 with lines
%    };
 \addplot[color=black,ultra thick,dotted,no marks]
table [col sep=space, x=Qy, y=Intensity]{Data/Gratings/simulatedBraggs3D.txt};
%    \addplot [no marks] gnuplot [raw gnuplot, color=black,ultra thick, dotted] {
%    set datafile separator " ";
%    plot "Data/Gratings/simulatedBraggs3D.txt" using
%    1:2 with lines
%    };
    \legend{\footnotesize Analytical ,\footnotesize Simulated}
    \end{semilogyaxis}
    \end{tikzpicture}
    \caption{3D line gratings.}
    \label{fig: 3DBraggs}
    \end{subfigure}
    \caption{Comparison of analytical and simulation line cut integration for 2D and 3D line gratings. The $q_x$ component of $\Vector{q}$ is at the location of the first order peak.}
    \label{fig:GratingComparison}
\end{figure}

 The second periodic structure test case is a set of parallel lines which form a grating structure. This type of morphology is observed in the directed self-assembly of block copolymers, or structures fabricated using lithographic processes; and is often seen in semiconducting manufacturing. We consider a single line grating morphology in 2D and 3D. The 3D morphology consists of single line grating extended in the Z direction. \figref{fig:Grating-setup} shows the setup for the grating simulation. The 2D morphology consists of $1024 \times 1024 \times 1$ voxels whereas the 3D morphology consists of $1024 \times 1024  \times 63$  voxels, with each voxel representing a physical dimension of $1 \times 1 \times 1$ nm$^3$. The simulation was carried out at 17 keV. The analytical results are calculated using a previously-published procedure \cite{sunday2015determining} in which the grating is discretized into a stack of trapezoids. The analytical solution for the Fourier Transform of a trapezoid is used to calculate the scattering intensity at each q position. \figref{fig:GratingComparison} compares the analytical and simulation results for the line gratings. The simulated results are in excellent agreement with the analytical results.

%  One case is an arrangement of circular domains on a 2D hexagonal lattice (representative of hexagonally-packed cylinders, a common block copolymer morphology, oriented parallel to the X-ray beam). The Cy-RSoXS simulation performed with rotation produces a scattering profile with peaks at the expected analytical locations ($q*, \sqrt{3}q*, \sqrt{4}q*, \sqrt{7}q*, \sqrt{9}q*,$ and so on) given the target lattice spacing of the input morphology. The second case is a set of parallel lines which form a grating-like structure. This morphology is observed in the directed self-assembly of block copolymers, or constructed using lithographic processes, and is important in semiconducting manufacturing.

\subsection{Orientation effect on polymer-grafted nanoparticles}
\begin{figure}[b!]
    \centering
    \includegraphics{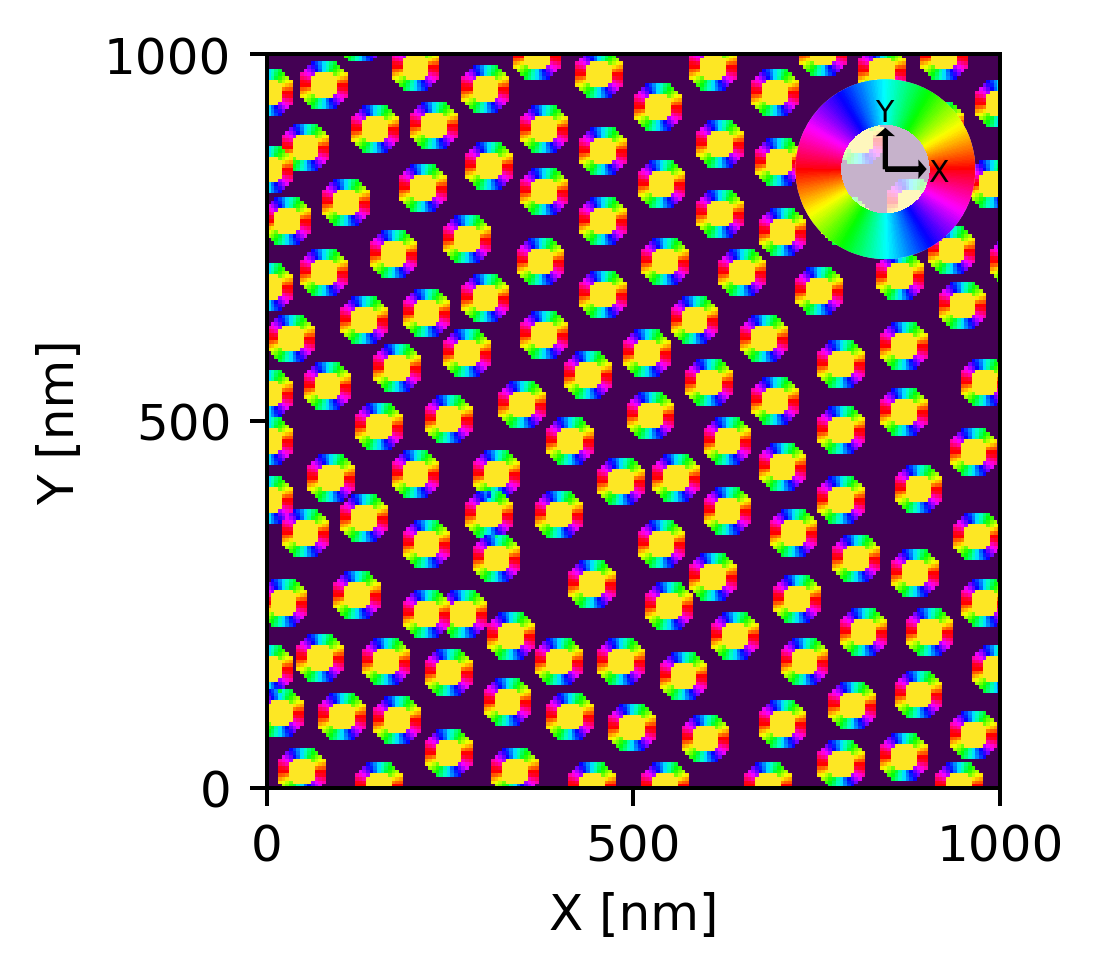}
    \caption{2D slice of 3D polymer-grafted nanoparticle (PGN) morphology. An oriented shell of polystyrene (PS) surrounds each gold nanoparticle core. The pixels in this image are colored by the values of the Euler angle $\phi^{PS})$, which exhibits a radial orientation relative to the particle centers. The orientation of the extraordinary axis of the dielectric function in real space relative to the $x$  and $y$ axes is shown in the inset color wheel. This 2D slice was collected near the particle equators such that $\phi^{PS}$~$\approx$~$\pi/2$ for all pixels.}
    \label{fig:PGNrealspace}
\end{figure}

\begin{figure*}[t!]
    \centering
    \includegraphics[width=0.99\linewidth]{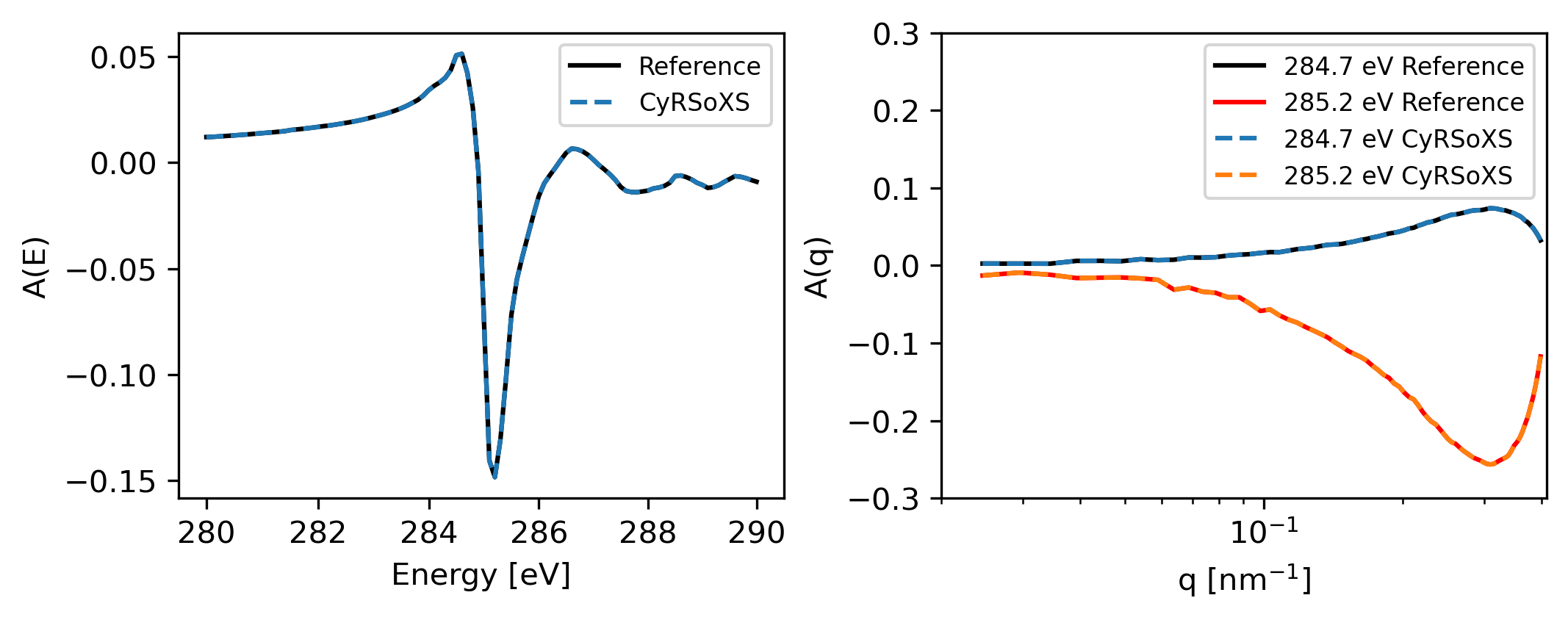}
    \caption{Scattering Anisotropy plotted versus select energies and q-values for the CyRSoXS and reference simulators}
    \label{fig:PGNtest}
\end{figure*}

All of the previous test cases dealt with isotropic materials. As the final validation case, we consider a film of polymer-grafted nanoparticles (PGNs)~\citep{mukherjee2021polarized}. Polystyrene chains are grafted onto gold nanoparticles, and the confinement of polystyrene chains near the nanoparticle surface results in radial stretching of the chains and a net molecular orientation. \figref{fig:PGNrealspace} is a 2D slice of the 3D morphology, showing the gold nanoparticle core surrounded by the oriented polystyrene shell, all embedded in a matrix of isotropic polystyrene. The CyRSoXS simulation is tested against the current state-of-the-art \prsoxs{} simulator \citep{gann2016origins}. \figref{fig:PGNtest} plots the scattering anisotropy at q~=~(0.02~to~0.04)~nm$^{-1}$ and two energies for the reference simulator and our GPU-accelerated \prsoxs{} simulator. Our implementation perfectly reproduces the results of the reference simulator.

\section{Performance}
\label{sec:Performance}
In this section, we report  the scaling of CyRSoXS with respect to variation in number of voxels and materials. All computation was carried out using \textsc{NVIDIA Voltas V-100} GPU with 32 GB of memory. 

\subsection{Performance with increasing number of voxels}

As a first scaling test, we considered performance with increase in number of voxels. The overall number of voxels  was varied from $128\times128\times16$ to $1024\times1024\times128$ with an increment of $2$ in each direction. \footnote{The $1024\times1024\times128$ voxel size is the largest size that fits into the memory of a 32 GB NVIDIA \textsc{V100} GPU.} The number of material is fixed to be 4 and the computation was carried out for 150 energy levels. For each energy level, Electric field $\Vector{e}$ was rotated from $0^o$ to $180^{o}$ at an increment of $2^o$.

\begin{figure}[t!]
\centering
\begin{subfigure}[b]{0.8\linewidth}
\begin{tikzpicture}
\begin{semilogyaxis}[
    width=\linewidth,
    height=0.38\linewidth,
    ybar,
    xticklabel style={rotate=0,font=\footnotesize},
    xticklabels = {
    % $64\times64\times8$,
    $128\times128\times16$,$256\times256\times32$,$512\times512\times64$,$1024\times1024\times128$}, % use the x column from the file for ticklabels
    xtick=data, % add a tick at every data point,
      enlarge x limits={abs=0.5}, % adjust space between axis edge and plot edge
    ylabel=Total time (s) $\rightarrow$,
    xlabel=Number of voxels $\rightarrow$,
    ]
    \addplot table[x expr=\coordindex,y=Total] {Data/Scaling/Alg1.txt}; 
\end{semilogyaxis}
\end{tikzpicture} 
\caption{Total time with variation in number of voxels.}
\label{fig:Alg1_time}
\end{subfigure}

\begin{subfigure}[b]{0.8\linewidth}
\vspace{5 mm}
\begin{tikzpicture}
\begin{axis}[
    width=\linewidth,
    height=0.38\linewidth,
    ybar stacked,
    ymin=0,
    xticklabel style={rotate=0,font=\footnotesize},
    xticklabels = {
    % $64\times64\times8$,
    $128\times128\times16$,$256\times256\times32$,$512\times512\times64$,$1024\times1024\times128$}, %
    % xticklabels from table={Data/data.txt}{x}, % use the x column from the file for ticklabels
    xtick=data, % add a tick at every data point,
     enlarge x limits={abs=0.5}, % adjust space between axis edge and plot edge
    ylabel=Percentage of time $\rightarrow$,
    xlabel=Number of voxels $\rightarrow$,
      legend style={at={(0.5,1.19)},anchor= north,legend columns=3}, 
    ]
    \addplot table[x expr=\coordindex,y expr=\thisrow{Polarization}*100/\thisrow{Sum}] {Data/Scaling/Alg1.txt}; 

     \addplot table[x expr=\coordindex,y expr=\thisrow{FFT}*100/\thisrow{Sum}] {Data/Scaling/Alg1.txt}; 
 
     \addplot table[x expr=\coordindex,y expr=(\thisrow{Scatter3DEwalds}+\thisrow{Malloc} +\thisrow{MemcopyCG} + \thisrow{Rotation} + \thisrow{MemcopyGC})*100/\thisrow{Sum}] {Data/Scaling/Alg1.txt}; 
     \legend{\footnotesize Polarization, \footnotesize FFT, \footnotesize Others}
\end{axis}
\end{tikzpicture}    
\caption{Percentage of time with variation in number of voxels.}
\label{fig:Alg1_percentage}
\end{subfigure}
\caption{Performance of ~\Algref{alg: prsoxs_comm_free} with variation in number of voxels. The number of material was fixed to 4. The time reported corresponds to computation of 150 energy level with $\Vector{e}$ rotated from $0^o$ to $180^o$ at increment of $2^o$ for each energy level.}
\label{fig:Alg1_performance}
\end{figure}

\figref{fig:Alg1_performance} compares the time with increase in the number of voxels for ~\Algref{alg: prsoxs_comm_free}. \figref{fig:Alg1_time} shows the variation of total wall-time with respect to the number of voxels. Overall we see a linear dependence ($\mathcal{O}(N)$), where $N$ is the total number of voxels. \figref{fig:Alg1_percentage} compares the percentage of time taken by different sections of the computation. The total time mostly is dominated by polarization computation (\eqnref{eq: polarization}) and FFT computation. The ``other" cost, which include Ewalds projection computation, image rotation and data transfer from CPU to GPU and vice--versa, form a significant fraction at lower resolution (i.e., smaller voxel sizes), but become insignificant at higher resolutions.

\begin{figure}[t!]
\centering
\begin{subfigure}[b]{0.8\linewidth}
\begin{tikzpicture}
\begin{semilogyaxis}[
    width=\linewidth,
    height=0.4\linewidth,
    ybar,
    xticklabel style={rotate=0,font=\footnotesize},
    xticklabels = {
    % $64\times64\times8$,
    $128\times128\times16$,$256\times256\times32$,$512\times512\times64$,$1024\times1024\times128$}, % use the x column from the file for ticklabels
    xtick=data, % add a tick at every data point,
    % enlarge x limits=0.03, % adjust space between axis edge and plot edge
    enlarge x limits={abs=0.5},
    ylabel=Total time (s) $\rightarrow$,
    xlabel=Number of voxels $\rightarrow$,
    ]
    \addplot table[x expr=\coordindex,y=Total] {Data/Scaling/Alg2.txt}; 
\end{semilogyaxis}
\end{tikzpicture}    
\caption{Total time with variation in number of voxels}
\label{fig:Alg2_time}
\end{subfigure}

\begin{subfigure}[b]{0.8\linewidth}
\vspace{5 mm}
\begin{tikzpicture}
\begin{axis}[
    width=\linewidth,
    height=0.4\linewidth,
    ybar stacked,
    ymin=0,
    xticklabel style={rotate=0,font=\footnotesize},
    xticklabels = {
    % $64\times64\times8$,
    $128\times128\times16$,$256\times256\times32$,$512\times512\times64$,$1024\times1024\times128$}, %
    % xticklabels from table={Data/data.txt}{x}, % use the x column from the file for ticklabels
    xtick=data, % add a tick at every data point,
    enlarge x limits={abs=0.5},
    ylabel=Percentage of time $\rightarrow$,
    xlabel=Number of voxels $\rightarrow$,
      legend style={at={(0.5,1.18)},anchor= north,legend columns=4}, 
    ]
     \addplot table[x expr=\coordindex,y expr=\thisrow{Nt}*100/\thisrow{Sum}] {Data/Scaling/Alg2.txt}; 
     
    \addplot table[x expr=\coordindex,y expr=\thisrow{Polarization}*100/\thisrow{Sum}] {Data/Scaling/Alg2.txt}; 

     \addplot table[x expr=\coordindex,y expr=\thisrow{FFT}*100/\thisrow{Sum}] {Data/Scaling/Alg2.txt}; 
    
     \addplot table[x expr=\coordindex,y expr=(\thisrow{Scatter3DEwalds}+\thisrow{Malloc} +\thisrow{MemcopyCG} + \thisrow{Rotation} + \thisrow{MemcopyGC})*100/\thisrow{Sum}] {Data/Scaling/Alg2.txt}; 
     \legend{\footnotesize Nt,\footnotesize Polarization,\footnotesize FFT,\footnotesize Others}
\end{axis}
\end{tikzpicture}
\caption{Percentage of time with variation in number of voxels}
\label{fig:Alg2_percentage}
\end{subfigure}
\caption{Performance of ~\Algref{alg: prsoxs} with variation in number of voxels. The number of material was fixed to 4. The time reported corresponds to computation of 150 energy level with $\Vector{e}$ rotated from $0^o$ to $180^o$ at increment of $2^o$ for each energy level.}
\label{fig:Alg2_performance}
\end{figure}
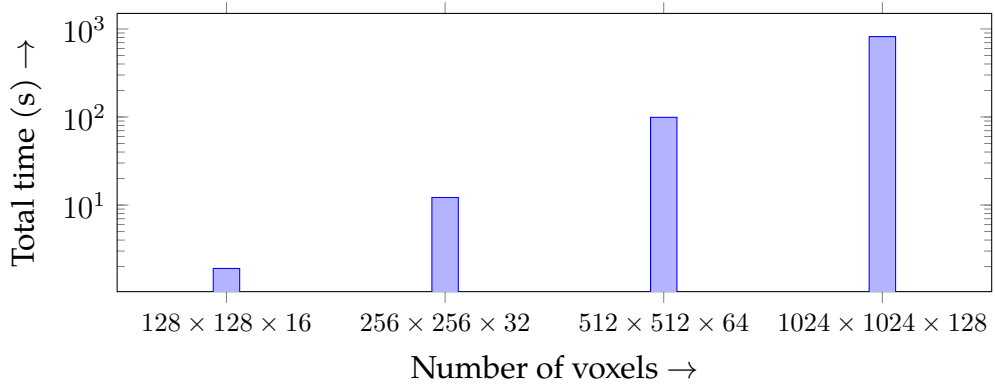
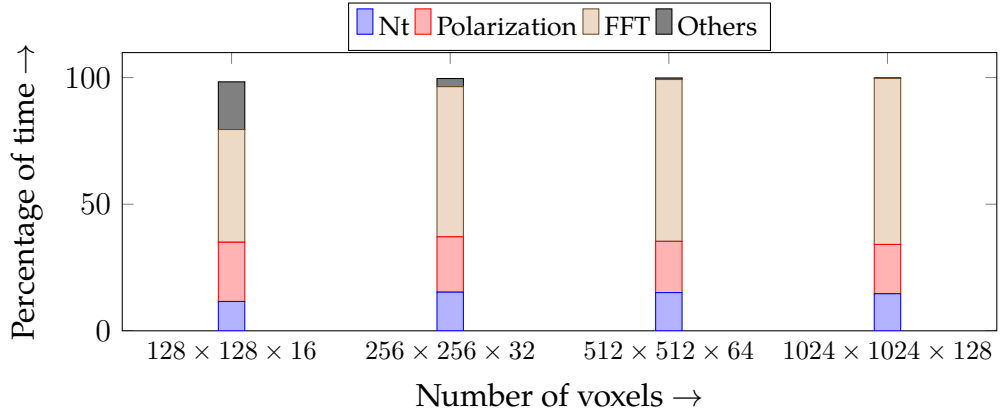

\figref{fig:Alg2_performance} compares the time with increase in the number of voxels for ~\Algref{alg: prsoxs}. \figref{fig:Alg2_time} shows the variation of total time whereas \figref{fig:Alg2_percentage} compares the percentage of time with increase in the number of voxels. We observe a similar performance behavior compared to the ~\Algref{alg: prsoxs_comm_free}, including $\mathcal{O}(N)$ scaling with increase in number of voxels.  The majority of the time is spent in computing $\Tensor{N_t}$ which also involves the copying data from CPU to GPU (\Algref{alg: localNr}), polarization computation and FFT computation.  The ``other" cost, similar to the previous algorithm, forms a significant chunk of percentage at lower resolution but becomes insignificant at higher resolutions.

\subsection{Performance with increasing number of materials: Communication minimization vs memory minimization algorithms}
As a next analysis, we compared the performance of both the algorithm with respect to increase in the number of materials. We considered a system with a voxel size of $2048\times2048\times64$. The computation was carried
out for 9 energy levels. For each energy level, Electric field $\Vector{e}$ was rotated from $0^o$ to $180^o$ at an
increment of $2^o$. We utilize 10 streams for the computation of ~\Algref{alg: prsoxs} to overlap computation and communication.

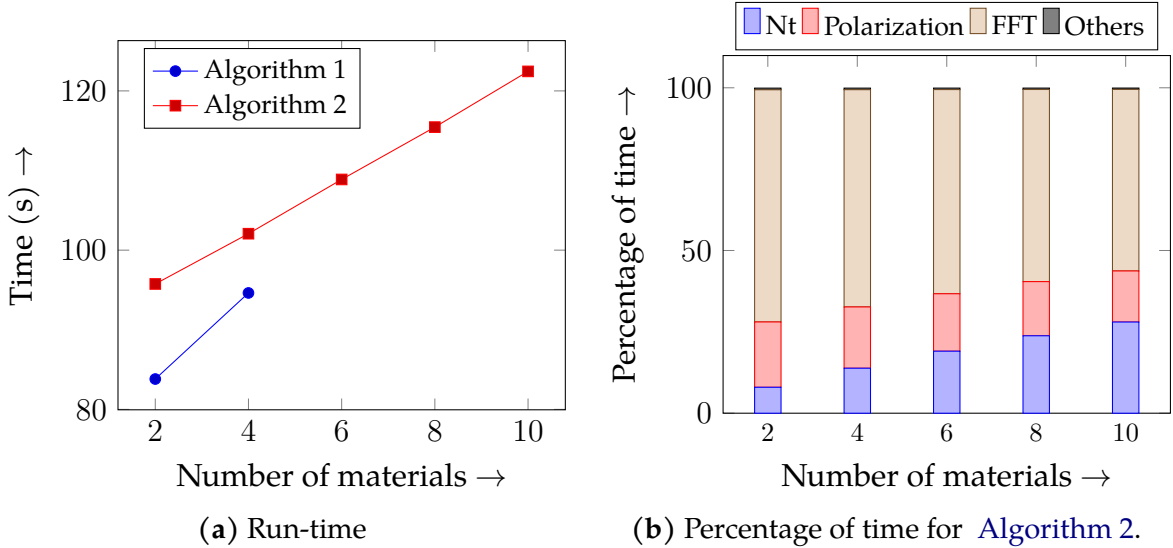
\begin{figure*}[t!]
    \centering
    \begin{subfigure}[b]{0.48\linewidth}
    \centering
        \begin{tikzpicture}
        \begin{axis}[
              %width=0.6\linewidth, % Scale the plot to \linewidth
              width=0.95\linewidth,
              %height=0.2\textwidth,
              %grid=major, % Display a grid
              %grid style={line width=.1pt, draw=gray!50}, % Set the style
              xlabel=Number of materials $\rightarrow$, % Set the labels
              ylabel=Time (s) $\rightarrow$,
            %   xmin=0,
            %   xmax=3.5,
            %   ymax = 1.1,
            %   legend style={at={(1.2,0.7)},anchor=north west,legend columns=1}, % Put the legend below the plot
              x tick label style={rotate=0,anchor=north}, % Display labels sideways
              legend style={at={(0.3,0.98)},anchor= north,legend columns=1}, % Put the legend below the plot
        ]
        \addplot
            table [col sep=space, x=NumMat, y=Total]{Data/Performance/Alg1.txt};
        \addplot 
            table [col sep=space, x=NumMat, y=Total]{Data/Performance/Alg2.txt};
        \legend{\footnotesize Algorithm 1 ,\footnotesize Algorithm 2}
        \end{axis}
    \end{tikzpicture}
    \caption{Run-time}
    \label{fig:mat_runtime}
    \end{subfigure}
    \vspace{0.02\linewidth}
    \begin{subfigure}[b]{0.48\linewidth}
    \centering
    \begin{tikzpicture}
        \begin{axis}[
        width=0.95\linewidth,
        height=0.8\linewidth,
        ybar stacked,
        ymin=0,
        xticklabel style={rotate=0,font=\footnotesize},
        xticklabels = {$2$,$4$,$6$,$8$,$10$}, %
        % xticklabels from table={Data/data.txt}{x}, % use the x column from the file for ticklabels
        xtick=data, % add a tick at every data point,
        % enlarge x limits=, % adjust space between axis edge and plot edge
        enlarge x limits={abs=0.5},
        ylabel=Percentage of time $\rightarrow$,
        xlabel=Number of materials $\rightarrow$,
        legend style={at={(0.5,1.15)},anchor= north,legend columns=4}, 
        ]
        \addplot table[x expr=\coordindex,y expr=\thisrow{Nt}*100/\thisrow{Sum}] {Data/Performance/Alg2.txt}; 
        
        \addplot table[x expr=\coordindex,y expr=\thisrow{Polarization}*100/\thisrow{Sum}] {Data/Performance/Alg2.txt}; 
        
        \addplot table[x expr=\coordindex,y expr=\thisrow{FFT}*100/\thisrow{Sum}] {Data/Performance/Alg2.txt}; 
        
        \addplot table[x expr=\coordindex,y expr=(\thisrow{Scatter3DEwalds}+\thisrow{Malloc} +\thisrow{MemcopyCG} + \thisrow{Rotation} + \thisrow{MemcopyGC})*100/\thisrow{Sum}] {Data/Performance/Alg2.txt}; 
        \legend{\footnotesize Nt, \footnotesize Polarization, \footnotesize FFT, \footnotesize Others}
        \end{axis}
        \end{tikzpicture}    
    \caption{Percentage of time for ~\Algref{alg: prsoxs}.}
    \label{fig:mat_percentage}
    \end{subfigure}
    \caption{Run time and percentage distribution of ~\prsoxs{} for $2048\times2048\times64$ morphology with increasing number of material for 9 different energy levels. $\Vector{e}$ rotated from $0^o$ to $180^o$ at increment of $2^o$ for each energy level. For ~\Algref{alg: prsoxs_comm_free}, the overall memory requirement exceeds the GPU memory size for number of material $>$ 4.}
    \label{fig:Performance_material}
\end{figure*}

\figref{fig:mat_runtime} compares the total time for both the algorithm. We see ~\Algref{alg: prsoxs_comm_free} to be faster than the ~\Algref{alg: prsoxs}. However, the overall slope, or the rate of increase in time with material size tends to the much steeper for ~\Algref{alg: prsoxs_comm_free} compared to the ~\Algref{alg: prsoxs}. This is because the polarization computation (\eqnref{eq: polarization}) involves a loop over the number of material. ~\Algref{alg: prsoxs_comm_free} performs this computation  for each rotation of $\Vector{e}$, whereas in case of ~\Algref{alg: prsoxs}, this computation is carried out once for each energy level and stored in $\Tensor{N_t}$. With increase in the number of material, this computation tends to dominate, and thus, we see a higher slope for ~\Algref{alg: prsoxs_comm_free}. Further, we observe that the memory requirement of the ~\Algref{alg: prsoxs_comm_free} exceeds the overall GPU memory for material size $> 4$, whereas ~\Algref{alg: prsoxs} continues to give a linear variation with increase in material size. This agrees with the memory requirement analysis in ~\secref{sec: Algorithm}. We recall that memory requirement of \Algref{alg: prsoxs_comm_free} exceeds \Algref{alg: prsoxs} for material size~$\geq$~3. 

\figref{fig:mat_percentage} shows the percentage of time for different sections of the ~\Algref{alg: prsoxs}. We see an increase in the time for $\Tensor{N_t}$ computation. This is expected as only $\Tensor{N_t}$ computation in ~\Algref{alg: prsoxs} depends on the number of materials. Overall, the time is mostly dominated by FFT computations.

\subsection{Scaling performance across multiple GPUs}
We parallelize the code with respect to the energy level across multiple GPUs. This makes the code embarrassingly parallel. Each GPU device allocates its own chunk of memory depending on the energy levels owned by it and performs the computation independently. We utilize \textsc{OpenMP} to schedule the threads with each thread handling a single GPU. This allows us to utilize all GPUs efficiently across a single node.

In order to demonstrate the scaling performance, we consider a server with 2 \textsc{NVIDIA V100} GPUs and analyzed the efficiency for different voxel sizes. We consider a material system with two different voxel size of $512 \times 512 \times 64$ and $1024\times1024\times 128$ and 4 materials. We consider 150 energy levels distributed across multiple GPUs. Overall, we see an ideal scaling behavior with both algorithms achieving $2\times$ speedup while utilizing 2 GPUs.

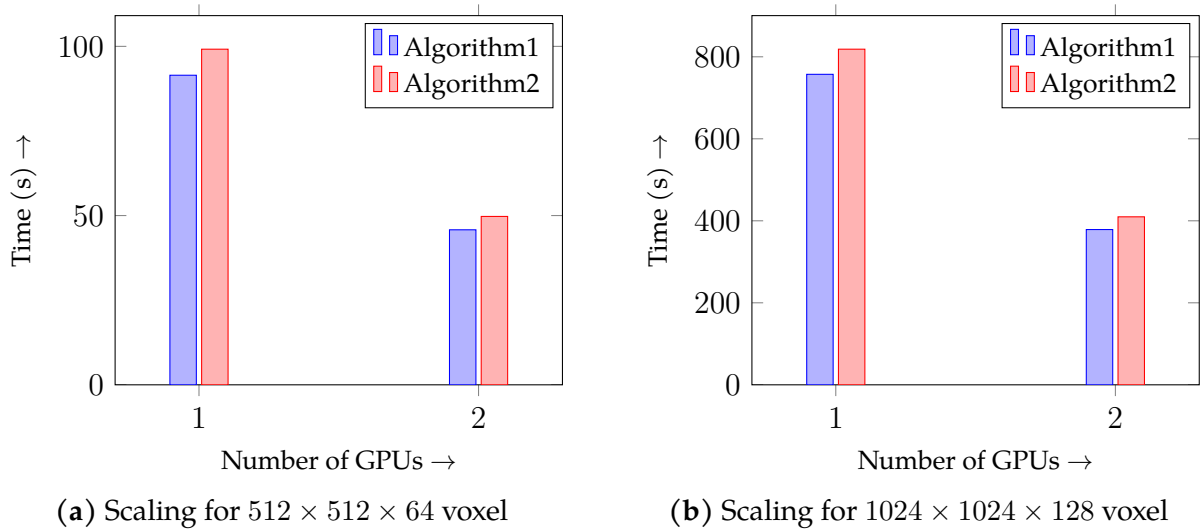
\begin{figure*}[h!]
\centering
\begin{subfigure}[b]{0.48\linewidth}
    \centering
    \begin{tikzpicture}
        \begin{axis}[
        width=0.95\linewidth, 
        ymin = 0,
        xlabel= \footnotesize Number of GPUs $\rightarrow$,
        ylabel= \footnotesize Time (s) $\rightarrow$,
        xmode=log,
        log ticks with fixed point,
        xtick=data,
        ybar, %added here
        enlarge x limits=0.3, % adjust space between axis edge and plot edge
        ]
        \addplot table[x=Thread,y=TimeAlg1, col sep=space] {Data/OpenMP/sys8.txt};
        \addplot table[x=Thread,y=TimeAlg2, col sep=space] {Data/OpenMP/sys8.txt};
        \legend{\footnotesize Algorithm1, \footnotesize Algorithm2}
        \end{axis}
    \end{tikzpicture}
    \caption{Scaling for $512\times512\times64$ voxel}
\end{subfigure}
\hspace{0.02\linewidth}
\begin{subfigure}[b]{0.48\linewidth}
    \centering
    \begin{tikzpicture}
        \begin{axis}[
        width=0.95\linewidth, 
        %   , height=8cm,
        %   grid=major,
        ymin = 0,
        xlabel= \footnotesize Number of GPUs $\rightarrow$,
        ylabel= \footnotesize Time (s) $\rightarrow$,
        xmode=log,
        log ticks with fixed point,
        xtick=data,
        ybar, %added here
        enlarge x limits=0.3, % adjust space between axis edge and plot edge
        ]
        \addplot table[x=Thread,y=TimeAlg1, col sep=space] {Data/OpenMP/sys16.txt};
        \addplot table[x=Thread,y=TimeAlg2, col sep=space] {Data/OpenMP/sys16.txt};
        \legend{\footnotesize Algorithm1, \footnotesize Algorithm2}
        \end{axis}
    \end{tikzpicture}
    \caption{Scaling for $1024\times1024\times128$ voxel}
\end{subfigure}
\caption{Scaling of ~\prsoxs{} simulator on multiple GPU. The number of material was fixed to 4. The
time reported corresponds to computation of 150 energy level distributed across multiple GPU with $\Vector{e}$ rotated from $0^o$ to $180^o$ at increment of $2^o$ for each energy level.}
\label{fig:Scaling}
\end{figure*}

\section{Python interface to CyRSoXS}
\label{sec:pybind}
In addition to GPU acceleration, we have added a python interface using Pybind11~\cite{jakob2019pybind11}. Pybind11 was designed to expose C++ data types to Python and vice-versa. One of the benefits of this approach is directly passing the morphology information via memory instead of performing file IO operations, which can be a major bottleneck for fitting and other inverse problems. Additionally, the output of the scattering pattern in the form of NumPy arrays enables users to use sophisticated python visualization libraries like Matplotlib~\cite{barrett2005matplotlib}, seaborn~\cite{waskom2020seaborn}, and develop Python-based post-processing tools. We also interface with the \texttt{cupy}~\cite{nishino2017cupy} library that enables morphology generation on GPU. A morphology generated on GPU can be directly passed to the simulator without copying data back and forth from the CPU. However, the morphology layout must strictly match the framework layout as shown in \figref{fig: memoryLayout} and described in \secref{sec:Morphology}.

We believe that the availability of the Python interface will give a major boost to inverse problems relating to the material design, as most of the Machine Learning (ML) or Data analysis (DA) toolkits \cite{garreta2013learning,chollet2018keras,abadi2016tensorflow,paszke2019pytorch} are currently Python-based. This interface will allow the users to seamlessly integrate their ML/DA models with the current framework.

\section{Conclusion}
\label{sec:Conclusion}
We have demonstrated a new \prsoxs{} virtual instrument with greatly increased performance compared to the state-of-the-art. Computations with this new virtual instrument are fast enough to enable practical data fitting by adjusting structure parameters using goal-seeking algorithms. The first fitting of orientational parameters to experimental \prsoxs~data was recently demonstrated using this virtual instrument to simulate polymer-grafted nanoparticles using a high-throughput multi-resolution parametric sweep of a 3-parameter system \citep{mukherjee2021polarized}. We have developed soon-to-be-published Python-driven workflows that demonstrate the practical use of this virtual instrument with other fitting methods, including genetic algorithm and Markov Chain Monte Carlo approaches. Close integration with Python environments affords opportunities to develop morphological models based on data fusion approaches, particularly leveraging real-space imaging, which reduces common questions of model uniqueness in fitting small-angle scattering data. The \prsoxs{} virtual instrument shows great promise as a cornerstone of future approaches for assimilating complementary data streams to construct complex and self-consistent material structure representations \textit{in silico} and ultimately to power inverse design frameworks that eliminate the need for costly Edisonian optimization approaches.

\section{Acknowledgements}
The authors acknowledge XSEDE grant number TG-CTS110007 for computing time and Iowa State University computing resources. KS and BG were supported in part by Office of Naval Research (ONR) Multiple University Research Initiative (MURI) 6119-ISU-ONR-2453. MLC and VGR acknowledge support by the National Science Foundation through Award No. DMR-1808622. VGR thanks the National Science Foundation Graduate Research Fellowship Program under Grant 1650114. 

% \section*{References}
% \bibliographystyle{elsarticle-num-names}

\bibliography{main.bib}

\providecommand{\latin}[1]{#1}
\makeatletter
\providecommand{\doi}
  {\begingroup\let\do\@makeother\dospecials
  \catcode`\{=1 \catcode`\}=2 \doi@aux}
\providecommand{\doi@aux}[1]{\endgroup\texttt{#1}}
\makeatother
\providecommand*\mcitethebibliography{\thebibliography}
\csname @ifundefined\endcsname{endmcitethebibliography}
  {\let\endmcitethebibliography\endthebibliography}{}
\begin{mcitethebibliography}{37}
\providecommand*\natexlab[1]{#1}
\providecommand*\mciteSetBstSublistMode[1]{}
\providecommand*\mciteSetBstMaxWidthForm[2]{}
\providecommand*\mciteBstWouldAddEndPuncttrue
  {\def\EndOfBibitem{\unskip.}}
\providecommand*\mciteBstWouldAddEndPunctfalse
  {\let\EndOfBibitem\relax}
\providecommand*\mciteSetBstMidEndSepPunct[3]{}
\providecommand*\mciteSetBstSublistLabelBeginEnd[3]{}
\providecommand*\EndOfBibitem{}
\mciteSetBstSublistMode{f}
\mciteSetBstMaxWidthForm{subitem}{(\alph{mcitesubitemcount})}
\mciteSetBstSublistLabelBeginEnd
  {\mcitemaxwidthsubitemform\space}
  {\relax}
  {\relax}

\bibitem[Wessels and Jayaraman(2021)Wessels, and
  Jayaraman]{wessels2021computational}
Wessels,~M.~G.; Jayaraman,~A. Computational Reverse-Engineering Analysis of
  Scattering Experiments (CREASE) on Amphiphilic Block Polymer Solutions:
  Cylindrical and Fibrillar Assembly. \emph{Macromolecules} \textbf{2021},
  \emph{54}, 783--796\relax
\mciteBstWouldAddEndPuncttrue
\mciteSetBstMidEndSepPunct{\mcitedefaultmidpunct}
{\mcitedefaultendpunct}{\mcitedefaultseppunct}\relax
\EndOfBibitem
\bibitem[Mukherjee \latin{et~al.}(2021)Mukherjee, Streit, Gann, Saurabh,
  Sunday, Krishnamurthy, Ganapathysubramanian, Richter, Vaia, and
  DeLongchamp]{mukherjee2021polarized}
Mukherjee,~S.; Streit,~J.~K.; Gann,~E.; Saurabh,~K.; Sunday,~D.~F.;
  Krishnamurthy,~A.; Ganapathysubramanian,~B.; Richter,~L.~J.; Vaia,~R.~A.;
  DeLongchamp,~D.~M. Polarized X-ray scattering measures molecular orientation
  in polymer-grafted nanoparticles. \emph{Nature Communications} \textbf{2021},
  \emph{12}, 1--10\relax
\mciteBstWouldAddEndPuncttrue
\mciteSetBstMidEndSepPunct{\mcitedefaultmidpunct}
{\mcitedefaultendpunct}{\mcitedefaultseppunct}\relax
\EndOfBibitem
\bibitem[Pryor \latin{et~al.}(2017)Pryor, Ophus, and Miao]{pryor2017streaming}
Pryor,~A.; Ophus,~C.; Miao,~J. A streaming multi-GPU implementation of image
  simulation algorithms for scanning transmission electron microscopy.
  \emph{Advanced structural and chemical imaging} \textbf{2017}, \emph{3},
  1--14\relax
\mciteBstWouldAddEndPuncttrue
\mciteSetBstMidEndSepPunct{\mcitedefaultmidpunct}
{\mcitedefaultendpunct}{\mcitedefaultseppunct}\relax
\EndOfBibitem
\bibitem[Reynolds \latin{et~al.}(2022)Reynolds, Callan, Saurabh, Murphy,
  Albanese, Chen, Wu, Gann, Hawker, Ganapathysubramanian, Bates, and
  Chabinyc]{reynolds2022simulation}
Reynolds,~V.~G.; Callan,~D.~H.; Saurabh,~K.; Murphy,~E.~A.; Albanese,~K.~R.;
  Chen,~Y.-Q.; Wu,~C.; Gann,~E.; Hawker,~C.~J.; Ganapathysubramanian,~B.;
  Bates,~C.~M.; Chabinyc,~M.~L. Simulation-guided analysis of resonant soft
  X-ray scattering for determining the microstructure of triblock copolymers.
  \emph{Molecular Systems Design \& Engineering} \textbf{2022}, \relax
\mciteBstWouldAddEndPunctfalse
\mciteSetBstMidEndSepPunct{\mcitedefaultmidpunct}
{}{\mcitedefaultseppunct}\relax
\EndOfBibitem
\bibitem[Axelrod \latin{et~al.}(2022)Axelrod, Schwalbe-Koda, Mohapatra,
  Damewood, Greenman, and G{\'o}mez-Bombarelli]{axelrod2022learning}
Axelrod,~S.; Schwalbe-Koda,~D.; Mohapatra,~S.; Damewood,~J.; Greenman,~K.~P.;
  G{\'o}mez-Bombarelli,~R. Learning matter: Materials design with machine
  learning and atomistic simulations. \emph{Accounts of Materials Research}
  \textbf{2022}, \emph{3}, 343--357\relax
\mciteBstWouldAddEndPuncttrue
\mciteSetBstMidEndSepPunct{\mcitedefaultmidpunct}
{\mcitedefaultendpunct}{\mcitedefaultseppunct}\relax
\EndOfBibitem
\bibitem[Vasudevan \latin{et~al.}(2021)Vasudevan, Pilania, and
  Balachandran]{vasudevan2021machine}
Vasudevan,~R.; Pilania,~G.; Balachandran,~P.~V. Machine learning for materials
  design and discovery. 2021\relax
\mciteBstWouldAddEndPuncttrue
\mciteSetBstMidEndSepPunct{\mcitedefaultmidpunct}
{\mcitedefaultendpunct}{\mcitedefaultseppunct}\relax
\EndOfBibitem
\bibitem[Guo \latin{et~al.}(2021)Guo, Yang, Yu, and Buehler]{guo2021artificial}
Guo,~K.; Yang,~Z.; Yu,~C.-H.; Buehler,~M.~J. Artificial intelligence and
  machine learning in design of mechanical materials. \emph{Materials Horizons}
  \textbf{2021}, \emph{8}, 1153--1172\relax
\mciteBstWouldAddEndPuncttrue
\mciteSetBstMidEndSepPunct{\mcitedefaultmidpunct}
{\mcitedefaultendpunct}{\mcitedefaultseppunct}\relax
\EndOfBibitem
\bibitem[Gomes \latin{et~al.}(2019)Gomes, Selman, and
  Gregoire]{gomes2019artificial}
Gomes,~C.~P.; Selman,~B.; Gregoire,~J.~M. Artificial intelligence for materials
  discovery. \emph{MRS Bulletin} \textbf{2019}, \emph{44}, 538--544\relax
\mciteBstWouldAddEndPuncttrue
\mciteSetBstMidEndSepPunct{\mcitedefaultmidpunct}
{\mcitedefaultendpunct}{\mcitedefaultseppunct}\relax
\EndOfBibitem
\bibitem[Collins and Gann(2022)Collins, and Gann]{collins2022resonant}
Collins,~B.~A.; Gann,~E. Resonant soft X-ray scattering in polymer science.
  \emph{Journal of Polymer Science} \textbf{2022}, \emph{60}, 1199--1243\relax
\mciteBstWouldAddEndPuncttrue
\mciteSetBstMidEndSepPunct{\mcitedefaultmidpunct}
{\mcitedefaultendpunct}{\mcitedefaultseppunct}\relax
\EndOfBibitem
\bibitem[Gann \latin{et~al.}(2016)Gann, Collins, Tang, Tumbleston, Mukherjee,
  and Ade]{gann2016origins}
Gann,~E.; Collins,~B.~A.; Tang,~M.; Tumbleston,~J.~R.; Mukherjee,~S.; Ade,~H.
  Origins of polarization-dependent anisotropic X-ray scattering from organic
  thin films. \emph{Journal of synchrotron radiation} \textbf{2016}, \emph{23},
  219--227\relax
\mciteBstWouldAddEndPuncttrue
\mciteSetBstMidEndSepPunct{\mcitedefaultmidpunct}
{\mcitedefaultendpunct}{\mcitedefaultseppunct}\relax
\EndOfBibitem
\bibitem[Jiao \latin{et~al.}(2017)Jiao, Ye, and Ade]{jiao2017quantitative}
Jiao,~X.; Ye,~L.; Ade,~H. Quantitative Morphology--Performance Correlations in
  Organic Solar Cells: Insights from Soft X-Ray Scattering. \emph{Advanced
  Energy Materials} \textbf{2017}, \emph{7}, 1700084\relax
\mciteBstWouldAddEndPuncttrue
\mciteSetBstMidEndSepPunct{\mcitedefaultmidpunct}
{\mcitedefaultendpunct}{\mcitedefaultseppunct}\relax
\EndOfBibitem
\bibitem[Ye \latin{et~al.}(2016)Ye, Jiao, Zhao, Zhang, Yao, Li, Ade, and
  Hou]{ye2016manipulation}
Ye,~L.; Jiao,~X.; Zhao,~W.; Zhang,~S.; Yao,~H.; Li,~S.; Ade,~H.; Hou,~J.
  Manipulation of domain purity and orientational ordering in high performance
  all-polymer solar cells. \emph{Chemistry of Materials} \textbf{2016},
  \emph{28}, 6178--6185\relax
\mciteBstWouldAddEndPuncttrue
\mciteSetBstMidEndSepPunct{\mcitedefaultmidpunct}
{\mcitedefaultendpunct}{\mcitedefaultseppunct}\relax
\EndOfBibitem
\bibitem[Song \latin{et~al.}(2018)Song, Gasparini, Nahid, Chen, Macphee, Zhang,
  Norman, Zhu, Bryant, and Ade]{song2018highly}
Song,~X.; Gasparini,~N.; Nahid,~M.~M.; Chen,~H.; Macphee,~S.~M.; Zhang,~W.;
  Norman,~V.; Zhu,~C.; Bryant,~D.; Ade,~H. A Highly Crystalline Fused-Ring
  n-Type Small Molecule for Non-Fullerene Acceptor Based Organic Solar Cells
  and Field-Effect Transistors. \emph{Advanced Functional Materials}
  \textbf{2018}, \emph{28}, 1802895\relax
\mciteBstWouldAddEndPuncttrue
\mciteSetBstMidEndSepPunct{\mcitedefaultmidpunct}
{\mcitedefaultendpunct}{\mcitedefaultseppunct}\relax
\EndOfBibitem
\bibitem[Song \latin{et~al.}(2019)Song, Gasparini, Nahid, Paleti, Li, Li, Ade,
  and Baran]{song2019efficient}
Song,~X.; Gasparini,~N.; Nahid,~M.~M.; Paleti,~S. H.~K.; Li,~C.; Li,~W.;
  Ade,~H.; Baran,~D. Efficient DPP Donor and Nonfullerene Acceptor Organic
  Solar Cells with High Photon-to-Current Ratio and Low Energetic Loss.
  \emph{Advanced Functional Materials} \textbf{2019}, \emph{29}, 1902441\relax
\mciteBstWouldAddEndPuncttrue
\mciteSetBstMidEndSepPunct{\mcitedefaultmidpunct}
{\mcitedefaultendpunct}{\mcitedefaultseppunct}\relax
\EndOfBibitem
\bibitem[Mukherjee \latin{et~al.}(2017)Mukherjee, Herzing, Zhao, Wu, Yu, Ade,
  DeLongchamp, and Richter]{mukherjee2017morphological}
Mukherjee,~S.; Herzing,~A.~A.; Zhao,~D.; Wu,~Q.; Yu,~L.; Ade,~H.;
  DeLongchamp,~D.~M.; Richter,~L.~J. Morphological characterization of
  fullerene and fullerene-free organic photovoltaics by combined real and
  reciprocal space techniques. \emph{Journal of Materials Research}
  \textbf{2017}, \emph{32}, 1921--1934\relax
\mciteBstWouldAddEndPuncttrue
\mciteSetBstMidEndSepPunct{\mcitedefaultmidpunct}
{\mcitedefaultendpunct}{\mcitedefaultseppunct}\relax
\EndOfBibitem
\bibitem[Litofsky \latin{et~al.}(2019)Litofsky, Lee, Aplan, Kuei, Hexemer,
  Wang, Wang, and Gomez]{litofsky2019polarized}
Litofsky,~J.~H.; Lee,~Y.; Aplan,~M.~P.; Kuei,~B.; Hexemer,~A.; Wang,~C.;
  Wang,~Q.; Gomez,~E.~D. Polarized soft X-ray scattering reveals chain
  orientation within nanoscale polymer domains. \emph{Macromolecules}
  \textbf{2019}, \emph{52}, 2803--2813\relax
\mciteBstWouldAddEndPuncttrue
\mciteSetBstMidEndSepPunct{\mcitedefaultmidpunct}
{\mcitedefaultendpunct}{\mcitedefaultseppunct}\relax
\EndOfBibitem
\bibitem[Attwood and Sakdinawat(2017)Attwood, and Sakdinawat]{attwood2017x}
Attwood,~D.; Sakdinawat,~A. \emph{X-rays and extreme ultraviolet radiation:
  principles and applications}; Cambridge university press, 2017\relax
\mciteBstWouldAddEndPuncttrue
\mciteSetBstMidEndSepPunct{\mcitedefaultmidpunct}
{\mcitedefaultendpunct}{\mcitedefaultseppunct}\relax
\EndOfBibitem
\bibitem[St{\"o}hr(1992)]{stohr1992nexafs}
St{\"o}hr,~J. \emph{NEXAFS spectroscopy}; Springer Science \& Business Media,
  1992; Vol.~25\relax
\mciteBstWouldAddEndPuncttrue
\mciteSetBstMidEndSepPunct{\mcitedefaultmidpunct}
{\mcitedefaultendpunct}{\mcitedefaultseppunct}\relax
\EndOfBibitem
\bibitem[Mannsfeld(2012)]{mannsfeld2012tune}
Mannsfeld,~S.~C. In tune with organic semiconductors. \emph{Nature materials}
  \textbf{2012}, \emph{11}, 489--490\relax
\mciteBstWouldAddEndPuncttrue
\mciteSetBstMidEndSepPunct{\mcitedefaultmidpunct}
{\mcitedefaultendpunct}{\mcitedefaultseppunct}\relax
\EndOfBibitem
\bibitem[Wang \latin{et~al.}(2010)Wang, Hexemer, Nasiatka, Chan, Young,
  Padmore, Schlotter, L{\"u}ning, Swaraj, and Watts]{wang2010resonant}
Wang,~C.; Hexemer,~A.; Nasiatka,~J.; Chan,~E.; Young,~A.; Padmore,~H.;
  Schlotter,~W.; L{\"u}ning,~J.; Swaraj,~S.; Watts,~B. Resonant soft X-ray
  scattering of polymers with a 2D detector: Initial results and system
  developments at the Advanced Light Source. IOP Conf. Ser.: Mater. Sci. Eng.
  2010; p 012016\relax
\mciteBstWouldAddEndPuncttrue
\mciteSetBstMidEndSepPunct{\mcitedefaultmidpunct}
{\mcitedefaultendpunct}{\mcitedefaultseppunct}\relax
\EndOfBibitem
\bibitem[Watts(2014)]{watts2014calculation}
Watts,~B. Calculation of the Kramers-Kronig transform of X-ray spectra by a
  piecewise Laurent polynomial method. \emph{Optics Express} \textbf{2014},
  \emph{22}, 23628--23639\relax
\mciteBstWouldAddEndPuncttrue
\mciteSetBstMidEndSepPunct{\mcitedefaultmidpunct}
{\mcitedefaultendpunct}{\mcitedefaultseppunct}\relax
\EndOfBibitem
\bibitem[Born and Wolf(2013)Born, and Wolf]{born2013principles}
Born,~M.; Wolf,~E. \emph{Principles of optics: electromagnetic theory of
  propagation, interference and diffraction of light}; Elsevier, 2013\relax
\mciteBstWouldAddEndPuncttrue
\mciteSetBstMidEndSepPunct{\mcitedefaultmidpunct}
{\mcitedefaultendpunct}{\mcitedefaultseppunct}\relax
\EndOfBibitem
\bibitem[Gann(2022)]{eliottGithub}
Gann,~E. Optical-Constants-Database. 2022;
  \url{https://github.com/EliotGann/Optical-Constants-Database}\relax
\mciteBstWouldAddEndPuncttrue
\mciteSetBstMidEndSepPunct{\mcitedefaultmidpunct}
{\mcitedefaultendpunct}{\mcitedefaultseppunct}\relax
\EndOfBibitem
\bibitem[Tatchev(2010)]{tatchev2010multiphase}
Tatchev,~D. Multiphase approximation for small-angle scattering. \emph{Journal
  of Applied Crystallography} \textbf{2010}, \emph{43}, 8--11\relax
\mciteBstWouldAddEndPuncttrue
\mciteSetBstMidEndSepPunct{\mcitedefaultmidpunct}
{\mcitedefaultendpunct}{\mcitedefaultseppunct}\relax
\EndOfBibitem
\bibitem[Wang \latin{et~al.}(2011)Wang, Lee, Hexemer, Kim, Zhao, Hasegawa, Ade,
  and Russell]{wang2011defining}
Wang,~C.; Lee,~D.~H.; Hexemer,~A.; Kim,~M.~I.; Zhao,~W.; Hasegawa,~H.; Ade,~H.;
  Russell,~T.~P. Defining the nanostructured morphology of triblock copolymers
  using resonant soft X-ray scattering. \emph{Nano letters} \textbf{2011},
  \emph{11}, 3906--3911\relax
\mciteBstWouldAddEndPuncttrue
\mciteSetBstMidEndSepPunct{\mcitedefaultmidpunct}
{\mcitedefaultendpunct}{\mcitedefaultseppunct}\relax
\EndOfBibitem
\bibitem[Virgili \latin{et~al.}(2007)Virgili, Tao, Kortright, Balsara, and
  Segalman]{virgili2007analysis}
Virgili,~J.~M.; Tao,~Y.; Kortright,~J.~B.; Balsara,~N.~P.; Segalman,~R.~A.
  Analysis of order formation in block copolymer thin films using resonant soft
  X-ray scattering. \emph{Macromolecules} \textbf{2007}, \emph{40},
  2092--2099\relax
\mciteBstWouldAddEndPuncttrue
\mciteSetBstMidEndSepPunct{\mcitedefaultmidpunct}
{\mcitedefaultendpunct}{\mcitedefaultseppunct}\relax
\EndOfBibitem
\bibitem[Freychet \latin{et~al.}(2018)Freychet, Cordova, McAfee, Kumar,
  Pandolfi, Anderson, Naulleau, Wang, and Hexemer]{freychet2018using}
Freychet,~G.; Cordova,~I.~A.; McAfee,~T.; Kumar,~D.; Pandolfi,~R.~J.;
  Anderson,~C.; Naulleau,~P.; Wang,~C.; Hexemer,~A. Using resonant soft x-ray
  scattering to image patterns on undeveloped resists. International Conference
  on Extreme Ultraviolet Lithography 2018. 2018; p 108090V\relax
\mciteBstWouldAddEndPuncttrue
\mciteSetBstMidEndSepPunct{\mcitedefaultmidpunct}
{\mcitedefaultendpunct}{\mcitedefaultseppunct}\relax
\EndOfBibitem
\bibitem[Sunday \latin{et~al.}(2015)Sunday, List, Chawla, and
  Kline]{sunday2015determining}
Sunday,~D.~F.; List,~S.; Chawla,~J.~S.; Kline,~R.~J. Determining the shape and
  periodicity of nanostructures using small-angle x-ray scattering.
  \emph{Journal of Applied Crystallography} \textbf{2015}, \emph{48},
  1355--1363\relax
\mciteBstWouldAddEndPuncttrue
\mciteSetBstMidEndSepPunct{\mcitedefaultmidpunct}
{\mcitedefaultendpunct}{\mcitedefaultseppunct}\relax
\EndOfBibitem
\bibitem[Jakob \latin{et~al.}(2019)Jakob, Rhinelander, and
  Moldovan]{jakob2019pybind11}
Jakob,~W.; Rhinelander,~J.; Moldovan,~D. pybind11--seamless operability between
  c++ 11 and python, 2017. \emph{URL https://github. com/pybind/pybind11}
  \textbf{2019}, \relax
\mciteBstWouldAddEndPunctfalse
\mciteSetBstMidEndSepPunct{\mcitedefaultmidpunct}
{}{\mcitedefaultseppunct}\relax
\EndOfBibitem
\bibitem[Barrett \latin{et~al.}(2005)Barrett, Hunter, Miller, Hsu, and
  Greenfield]{barrett2005matplotlib}
Barrett,~P.; Hunter,~J.; Miller,~J.~T.; Hsu,~J.-C.; Greenfield,~P.
  matplotlib--A Portable Python Plotting Package. Astronomical data analysis
  software and systems XIV. 2005; p~91\relax
\mciteBstWouldAddEndPuncttrue
\mciteSetBstMidEndSepPunct{\mcitedefaultmidpunct}
{\mcitedefaultendpunct}{\mcitedefaultseppunct}\relax
\EndOfBibitem
\bibitem[Waskom \latin{et~al.}(2020)Waskom, Botvinnik, Gelbart, Ostblom,
  Hobson, Lukauskas, Gemperline, Augspurger, Halchenko, and
  Warmenhoven]{waskom2020seaborn}
Waskom,~M.; Botvinnik,~O.; Gelbart,~M.; Ostblom,~J.; Hobson,~P.; Lukauskas,~S.;
  Gemperline,~D.~C.; Augspurger,~T.; Halchenko,~Y.; Warmenhoven,~J. Seaborn:
  statistical data visualization. \emph{Astrophysics Source Code Library}
  \textbf{2020}, ascl--2012\relax
\mciteBstWouldAddEndPuncttrue
\mciteSetBstMidEndSepPunct{\mcitedefaultmidpunct}
{\mcitedefaultendpunct}{\mcitedefaultseppunct}\relax
\EndOfBibitem
\bibitem[Nishino and Loomis(2017)Nishino, and Loomis]{nishino2017cupy}
Nishino,~R.; Loomis,~S. H.~C. CuPy: A NumPy-compatible library for NVIDIA GPU
  calculations. \emph{31st confernce on neural information processing systems}
  \textbf{2017}, 151\relax
\mciteBstWouldAddEndPuncttrue
\mciteSetBstMidEndSepPunct{\mcitedefaultmidpunct}
{\mcitedefaultendpunct}{\mcitedefaultseppunct}\relax
\EndOfBibitem
\bibitem[Garreta and Moncecchi(2013)Garreta, and
  Moncecchi]{garreta2013learning}
Garreta,~R.; Moncecchi,~G. \emph{Learning scikit-learn: machine learning in
  python}; Packt Publishing Ltd, 2013\relax
\mciteBstWouldAddEndPuncttrue
\mciteSetBstMidEndSepPunct{\mcitedefaultmidpunct}
{\mcitedefaultendpunct}{\mcitedefaultseppunct}\relax
\EndOfBibitem
\bibitem[Chollet(2018)]{chollet2018keras}
Chollet,~F. Keras: The python deep learning library. \emph{Astrophysics Source
  Code Library} \textbf{2018}, ascl--1806\relax
\mciteBstWouldAddEndPuncttrue
\mciteSetBstMidEndSepPunct{\mcitedefaultmidpunct}
{\mcitedefaultendpunct}{\mcitedefaultseppunct}\relax
\EndOfBibitem
\bibitem[Abadi \latin{et~al.}(2016)Abadi, Barham, Chen, Chen, Davis, Dean,
  Devin, Ghemawat, Irving, and Isard]{abadi2016tensorflow}
Abadi,~M.; Barham,~P.; Chen,~J.; Chen,~Z.; Davis,~A.; Dean,~J.; Devin,~M.;
  Ghemawat,~S.; Irving,~G.; Isard,~M. Tensorflow: A system for large-scale
  machine learning. 12th $\{$USENIX$\}$ symposium on operating systems design
  and implementation ($\{$OSDI$\}$ 16). 2016; pp 265--283\relax
\mciteBstWouldAddEndPuncttrue
\mciteSetBstMidEndSepPunct{\mcitedefaultmidpunct}
{\mcitedefaultendpunct}{\mcitedefaultseppunct}\relax
\EndOfBibitem
\bibitem[Paszke \latin{et~al.}(2019)Paszke, Gross, Massa, Lerer, Bradbury,
  Chanan, Killeen, Lin, Gimelshein, and Antiga]{paszke2019pytorch}
Paszke,~A.; Gross,~S.; Massa,~F.; Lerer,~A.; Bradbury,~J.; Chanan,~G.;
  Killeen,~T.; Lin,~Z.; Gimelshein,~N.; Antiga,~L. Pytorch: An imperative
  style, high-performance deep learning library. \emph{arXiv preprint
  arXiv:1912.01703} \textbf{2019}, \relax
\mciteBstWouldAddEndPunctfalse
\mciteSetBstMidEndSepPunct{\mcitedefaultmidpunct}
{}{\mcitedefaultseppunct}\relax
\EndOfBibitem
\end{mcitethebibliography}

% \appendix
% \input{appendix}
\end{document}